\documentclass[12pt]{iopart}

\usepackage{iopams}  
\usepackage{graphicx} 
\usepackage{subfig}
\usepackage{multirow}
\usepackage{dsfont}
\usepackage{cite}
\usepackage[dvipsnames]{xcolor}
\newcommand{\Pf}{\mathrm{Pf}}
\newcommand{\E}{\mathrm{E}}

\begin{document}

\title[Bures--Hall Ensemble: Spectral Densities and Average Entropies]{Bures--Hall Ensemble: Spectral Densities and Average Entropies }

\author{Ayana Sarkar* \& Santosh Kumar$^{\dagger}$ }

\address{Department of Physics, Shiv Nadar University, Gautam Buddha Nagar, Uttar Pradesh -- 201314, India}
\ead{*ayana1994s@gmail.com, $^{\dagger}$skumar.physics@gmail.com}
\vspace{10pt}

\begin{abstract}
We consider an ensemble of random density matrices distributed according to the Bures measure. The corresponding joint probability density of eigenvalues is described by the fixed trace Bures--Hall ensemble of random matrices which, in turn, is related to its unrestricted trace counterpart via a Laplace transform. We investigate the spectral statistics of both these ensembles and, in particular, focus on the level density, for which we obtain exact closed-form results involving Pfaffians. In the fixed trace case, the level density expression is used to obtain an exact result for the average Havrda--Charv\'at--Tsallis (HCT) entropy as a finite sum. Averages of von Neumann entropy, linear entropy and purity follow by considering appropriate limits in the average HCT expression. Based on exact evaluations of the average von Neumann entropy and the average purity, we also conjecture very simple formulae for these, which are similar to those in the Hilbert--Schmidt ensemble.
\end{abstract}

%
\vspace{2pc}

\section{Introduction}
\label{SecInt}

The density matrix formalism was introduced by von Neumann to describe statistical concepts in quantum mechanics~\cite{Neumann1927}. It plays a fundamental role in quantum mechanics and provides a natural approach to deal with mixed states~\cite{Haar1961,BZ2006}. Given the set of finite-size density matrices, it is now well acknowledged that, there is no unique measure which can be used to describe it~\cite{BZ2006,ZS2001,Slater1999,BS2001,SZ2003,ZS2003,SZ2004}. Therefore, one seeks a \emph{good} and useful measure which can be associated with these density matrices and consequently with the corresponding eigenvalues~\cite{BZ2006,ZS2001,Slater1999,BS2001,SZ2003,ZS2003,SZ2004}.

One of the ways to induce a measure over the space of random density matrices is via the operation of partial tracing~\cite{Brausntein1996,Hall1998}. This relates to the idea of \emph{purification} in which a mixed state $\rho$, acting on an $n$-dimensional Hilbert space $\mathcal{H}_n$, can be viewed as a reduced state obtained by partial tracing a pure state belonging to a composite Hilbert space $\mathcal{H}_n\otimes\mathcal{H}_m$~\cite{NC2000}. The auxiliary subsystem associated with $\mathcal{H}_m$ may be interpreted as the environment. In this approach, along with the dimension $n$ of the given density matrices, one ends up with an additional parameter at hand, viz. $m$ associated with the auxiliary subsystem. To elaborate, one considers a random pure state $|\varphi \rangle \in \mathcal{H}_{n} \otimes \mathcal{H}_{m}$ of a composite bipartite system of size $n \times m$ with $m \geq n$. Upon partial tracing over the $m$-dimensional environment, one obtains a reduced state of size $n$. The corresponding reduced density matrix is given by 
\begin{equation}
\rho = \frac{\tr_m(|\varphi\rangle\langle\varphi|)}{\langle \varphi|\varphi \rangle},
\end{equation}
and gives rise to the Hilbert-Schmidt measure,
\begin{equation}
\label{PHS}
P_\mathrm{HS}(\rho)\propto \,\delta(\tr\rho-1)(\det \rho)^{m-n}\,\Theta(\rho).
\end{equation}
Here $\Theta(\rho)$ is the Heaviside theta function with matrix argument, which implies the positive-definiteness of $\rho$. It is of interest to discuss the limiting situations for the Hilbert-space dimension $m$ of the auxiliary subsystem. For $m=1$, in view of our assumption $n\le m$, $n$ also takes the value of 1. As a consequence, the density matrix in this case is one-dimensional and has element 1. However, if we do let $n>m$, then $\rho$ will be $n$-dimensional and in its eigenbasis it will have one of the diagonal entries as 1 and rest $n-1$ as 0s. Physically, in the $m=1$ case, $\rho$ represents a pure state. The other extreme is obtained for $m\to\infty$. In this limit, as argued by Hall~\cite{Hall1998}, the distribution $P_\mathrm{HS}(\rho)$ tends to become a delta function, $\delta(\rho-n^{-1}\mathds{1}_n)$. All the $n$ eigenvalues of $\rho$ therefore assume the value $1/n$. The reduced density matrix $\rho$ in this case signifies a maximally mixed state.
For the Bures-Hall measure, one considers a symmetric superposition of two pure states of the composite system, one of which is a local unitary transformed copy of the other~\cite{BZ2006,OSZ2010,ZPNC2011}, viz., $|\psi \rangle\sim |\varphi \rangle+(U\otimes \mathds{1}_m)|\varphi \rangle$. Here $U$ is an $n$-dimensional unitary matrix taken from the measure $|\mathrm{det}(\mathds{1}_n + U)|^{2(m-n)}d\mu(U_n)$, with $d\mu(U_n)$ being the Haar measure. The reduced density matrix obtained in this case,
\begin{equation}
\label{PBHM}
\nonumber
\rho = \frac{\tr_m(|\psi\rangle\langle\psi|)}{\langle \psi|\psi \rangle},
\end{equation}
belongs to the Bures-Hall measure described by
\begin{eqnarray}
\label{PBH}
\nonumber
P_\mathrm{BH}(\rho)&\propto& \,\delta(\tr\rho-1)(\det \rho)^{m-n}\,\Theta(\rho)\int d[X] e^{-\tr \rho X^2}\\
&\propto &\,\delta(\tr\rho-1)\frac{(\det \rho)^{m-n-1/2}}{\prod_{j>k}{(\mu_j+\mu_k})}\,\Theta(\rho).
\end{eqnarray}
In the first line of the above equation, $X$ is a random Hermitian matrix and $d[X]$ is the corresponding flat measure, i.e., product of differential of independent components in $X$. In the second line $\{\mu_j\}\in[0,1]$ are the eigenvalues of the density matrix $\rho$. In both the limits $m=1$ and $m\to\infty$, the behavior of $\rho$ in the Bures-Hall case is identical to that in the Hilbert-Schmidt case.

Another approach for assigning a measure over random density matrices is to consider certain distance metrics on the space of mixed states. The two popular and physically relevant choices are the Hilbert--Schmidt distance~\cite{SZ2004,BZ2006} and the Bures distance~\cite{Bures1969,Uhlmann1976,Hubner1992}. 
The Hilbert--Schmidt distance between two density matrices $\rho_1$ and $\rho_2$ is defined as the Frobenius norm of their difference, i.e.,
\begin{equation}
D_\mathrm{HS}(\rho_1,\rho_2)=\sqrt{\tr[(\rho_1-\rho_2)^2]}.
\end{equation}
 The Bures distance, on the other hand, is given by~\cite{Bures1969}
\begin{equation}
D_{B} (\rho_{1},\rho_{2}) = \sqrt{2 - 2\tr(\sqrt{\rho_1}\rho_2\sqrt{\rho_1})^{1/2}}.
\end{equation}
It should be noted that the Bures distance is a function of \emph{fidelity}~\cite{Jos1994}, $F(\rho_{1},\rho_{2})=[\tr(\sqrt{\rho_{1}}\rho_{2}\sqrt{\rho_{1}})^{1/2}]^2$, which is a commonly used quantity in the field of quantum information~\cite{NC2000}. Fidelity allows one to judge the \emph{proximity} of a pair of mixed states and reduces to unity if the two states are identical. The Hilbert-Schmidt metric is Riemannian but not monotone, while the Bures metric is both Riemannian and monotone~\cite{BZ2006}. The monotonicity of the Bures metric guarantees that it does not grow under the action of a stochastic map, i.e. a completely positive trace preserving map~\cite{SZ2003}. Being a function of fidelity, the Bures metric is also Fubini-Study adjusted, i.e. for pure states it agrees with the natural geometry on them. Additionally, the Bures distance, in the subspace of diagonal matrices induces the statistical distance which is the \emph{Fisher-Rao} metric~\cite{Fisher1925}. It turns out that the Bures metric is the only monotone metric which is simultaneously Fisher adjusted and Fubini-Study adjusted~\cite{SZ2003}. These outstanding mathematical properties of the Bures metric provides additional inspiration to study the geometry it induces in the space of mixed quantum states. Interestingly, the measure induced by the above two distance metrics coincide with~(\ref{PHS}) and~(\ref{PBH}) if $m$ is set equal to $n$, i.e., if the Hilbert-space dimension of the environment is same as that of the system under observation. We should remark here that the terms Hilbert-Schmidt measure and Bures-Hall measure are conventionally used for this $m=n$ case, and therefore~(\ref{PHS}) and~(\ref{PBH}) can be seen as generalizations.

The Hilbert--Schmidt measure is commonly known as the fixed trace Wishart-Laguerre ensemble in the context of random matrix theory and has been extensively studied. Consequently, the corresponding spectral statistics and behavior of the associated observables are fairly well understood. For instance, based on the knowledge of the joint probability density (jpd) of eigenvalues~\cite{Lubkin1978,LP1988,Page1993,Hall1998,BS2001,ZS2001}, we know explicit answers for the level density and two-point correlation function, moments and cumulants of the eigenvalues and the entropy measures, asymptotic and universal behavior and also extreme eigenvalue distributions and moments~\cite{SZ2004,KAT2008,ATK2009,Vivo2010,KP2011,Page1993,Sen1996,ZS2003,CSZ2006,Giraud2007,BTL2012,BN2012,VPO2016,Wei2017,Wei2018,MBL2008,CLZ2010,AV2011,LZ2011,FMPPS2008,PFPPS2010,NMV2010,NMV2011,Majumdar2011,KSA2017,TLSB2018}. In comparison, the fixed trace Bures--Hall ensemble has been explored very little due to its more involved mathematical structure. The jpd of eigenvalues for this ensemble was derived by Hall in~\cite{Hall1998}. \.Zyczkowski, Sommers and co-workers have obtained several key results pertaining to the fixed trace Bures--Hall ensemble in the study of statistical distribution of random density matrices and the associated entropy measures~\cite{ZS2001,SZ2003,SZ2004,BZ2006,OSZ2010,ZPNC2011}. Borot and Nadal have obtained the purity distribution for a generalized version of the Bures--Hall fixed trace ensemble in the large dimension limit~\cite{BN2012}. The corresponding unrestricted trace variant has been investigated by Forrester and Kieburg in connection with the Cauchy two-matrix model~\cite{FK2016}. This connection was discovered by Bertola {\it et al.} while investigating the Cauchy two-matrix model~\cite{BGS2009}. Despite these invaluable contributions, there are several aspects related to the Bures--Hall ensemble that remain to be explored.

In this work, we investigate the spectral statistics of both unrestricted trace and fixed trace variants of the Bures--Hall ensemble. For the former, we obtain an exact result for the $r$-point correlation function of arbitrary order in terms of a Pfaffian. This Pfaffian expression offers an alternative representation for the correlation function than the one derived by Forrester and Kieburg~\cite{FK2016}. We then focus on the level density and use it to obtain the corresponding exact closed-form expression for the fixed trace ensemble. This, in turn, is used to calculate the average Havrda--Charv\'at--Tsallis (HCT) entropy~\cite{HC1967,Tsallis1988} of random density matrices which are described by the fixed trace Bures--Hall ensemble. Appropriate limits of the HCT entropy also lead to exact expressions for the average von-Neumann entropy and the average linear entropy or, equivalently, the average purity. Based on exact evaluations, we also conjecture very simple formulae for the average von-Neumann entropy and the average purity. Finally, we validate these analytical results using numerical simulation based on Dyson's log-gas formalism~\cite{Dyson1972,Mehta2004,Forrester2010}.

The presentation scheme of the paper is as follows. In section~\ref{SecUTBHE} we derive exact results pertaining to the spectral statistics of unrestricted trace Bures--Hall ensemble. This is then used in section~\ref{SecFTBHE} to obtain an exact result for the level density in the fixed trace case. In section~\ref{SecEM} we derive exact expressions for the average entropies. In section~\ref{SecSum} we conclude with a brief summary of our results and also indicate directions in which this work can be extended. Appendices collect details of the derivations of the analytical results presented in this paper.

\section{\label{sec:level3} Unrestricted trace Bures--Hall ensemble}
\label{SecUTBHE}

We begin with the unrestricted trace Bures--Hall ensemble. The matrices constituting this ensemble are given by~\cite{FK2016,ZPNC2011,OSZ2010}
\begin{equation}
\label{matrixB}
B=(\mathds{1}_n+U)G G^\dag (\mathds{1}_n+U^\dag).
\end{equation}
In this, with $n\leq m$, $G$ is an $n\times m$-dimensional complex Ginibre random matrix having the associated probability measure 
\begin{equation}
\label{PGG}
P_G(G)dG\propto\exp(-v^2\tr GG^\dag)dG
\end{equation}
 with $v^2=4$, and $U$ is an $n\times n$-dimensional random unitary matrix from the measure 
 \begin{equation}
 \label{PUU}
 P_U(U)d\mu(U_n)\propto |\mathrm{det}(\mathds{1}_n + U)|^{2(m-n)}d\mu(U_n).
 \end{equation}
 Here $dG$ represents the flat measure given by the product of differentials of all independent components in $G$, and  $d\mu(U_n)$ is the Haar measure on the group of $n$-dimensional unitary matrices. The related $m\times m$ random matrix 
\begin{equation}
\label{matrixBp}
B'=G^\dag (\mathds{1}_n+U^\dag)(\mathds{1}_n+U)G
\end{equation}
possesses $n$ eigenvalues identical to $B$ and, in addition, has $m-n$ generic zero eigenvalues.

The jpd of eigenvalues $(\lambda_{j} \, \epsilon \, [0, \infty), j = 1....n)$ for the random matrix $B$ is given by~\cite{ZS2001,FK2016},  
\begin{eqnarray}
\label{jpd-unres}
\mathcal{P}(\lambda_1,\dots,\lambda_n)=C\,\frac{\Delta^2(\{\lambda\})}{\Delta_{+}(\{\lambda\})}\prod_{i=1}^{n}\lambda_{i}^{\alpha} e^{-\lambda_i},
\end{eqnarray}
where, $C$ is the normalization factor, $\Delta(\{\lambda\})=\det[\lambda_k^{j-1}]=\prod_{j<k}(\lambda_k-\lambda_j)$ is the Vandermonde determinant, $\Delta_{+}(\{\lambda\})=\prod_{j<k}(\lambda_k+\lambda_j)$,
and $\alpha=m-n-1/2$, which assumes half-integer values only. However, one may relax this parameter to $\alpha>-1$ if the above jpd is defined without reference to the matrix model~(\ref{matrixB}). We should remark that~(\ref{jpd-unres}) pertains to a generalized Bures--Hall ensemble since the standard one corresponds to the case $m=n$, or equivalently $\alpha=-1/2$.
The jpd in~(\ref{jpd-unres}) is related to that of the fixed trace Bures--Hall ensemble of random density matrices via a Laplace transform~\cite{FK2016}, as discussed in section~\ref{SecFTBHE}. Interestingly, the above jpd also connects to the $O(1)$ matrix model~\cite{Kostov1989,EK1995}, as was discovered by Bertola {\it et al.} while investigating the Cauchy two-matrix model~\cite{BGS2009}. Later on, Forrester and Kieburg demonstrated the explicit relationship between the unrestricted Bures--Hall ensembles and the Cauchy two-matrix model in~\cite{FK2016}. This is very interesting since the former constitutes a Pfaffian point process, while the latter corresponds to a determinantal point process. Also, very recently, Muttalib--Borodin kind of deformation has been considered in the Cauchy two-matrix model and the unrestricted Bures--Hall ensemble by Forrester and Li~\cite{FL2018}. Furthermore, Hu and Li have shown that the partition function of the unrestricted trace Bures-Hall ensemble can be identified as the $\tau$-function of BKP and DKP hierarchies~\cite{HL2017}.

To proceed, we need to rewrite the jpd in~(\ref{jpd-unres}) as a product of a determinant and a Pfaffian. The Pfaffian for a $2n \times 2n$ antisymmetric matrix $A$ is defined as~\cite{Mehta2004,Forrester2010},
\begin{equation}
\label{Pfdef}
\Pf[A] = \sum\limits_{\mathrm{p}} \sigma_{\mathrm{p}} A_{i_{1},i_{2}} A_{i_{3},i_{4}} ...A_{i_{2n-1},i_{2n}}.
\end{equation}
The sum in~(\ref{Pfdef}) is over all possible permutations, 
\begin{eqnarray}
\mathrm{p} = \left(\begin{array}{cccc}
1 & 2 &\cdots & 2n \\
 i_{1} & i_{2} & \cdots & i_{2n}
\end{array}\right), 
\end{eqnarray}
with restrictions that $i_{1} < i_{2}, i_{3} < i_{4},...., i_{2n-1} < i_{2n} ; i_{1} < i_{3} < i_{2n-1}$ and $\sigma_{p}$ is the sign of the permutation. Also the pfaffian is associated to the determinant as 
\begin{equation}
\det[A] =\left(\Pf[A]\right)^{2}.
\end{equation}
In connection with the random matrix theory, Pfaffian (or equivalently quaternion-determinant) based formulae were introduced by Dyson to write down the eigenvalue correlation functions for circular orthogonal and symplectic ensembles~\cite{Dyson1970}. Pfaffian based techniques and results now constitute an indispensable part of random matrix theory and have been applied in several contexts~\cite{Mehta2004, Forrester2010,Mehta1971,Dyson1972,Mehta1976,BN1991,Nagao2007,AK2007,KG2010}. One of the most noteworthy applications of Pfaffians is in the random matrix ensembles modeling crossovers between various symmetry classes~\cite{PM1983,MP1983,PS1991,FNH1999,NF2003,KP2009,KP2011b}. 
Presently, for the jpd~(\ref{jpd-unres}), we use Schur's Pfaffian identity~\cite{Schur1911, IOTZ1995, FK2016},
\begin{equation}
\label{SchurPf}
\fl
\prod_{1\le j<k\le n}\frac{x_k-x_j}{x_k+x_j}=\cases{
\Pf\left[(x_k-x_j)/(x_k+x_j)\right]_{j,k=1,...,n}, & $n$  even,\\
\Pf\left[\begin{array}{cc}
\left[(x_k-x_j)/(x_k+x_j)\right]_{j,k=1,...,n} & \left[1\right]_{j=1,...,n} \\
\left[-1\right]_{k=1,...,n} & 0
\end{array}\right], & $n$  odd,
}
\end{equation}
and the result $\Delta(\{\lambda\})\prod_{i=1}^{n}\lambda_{i}^{\alpha} e^{-\lambda_i}=\det[\lambda_k^{j+\alpha-1} e^{-\lambda_{k}}]$. As a consequence, we have
\begin{eqnarray}
\mathcal{P}(\lambda_1,...,\lambda_n)=C \, \det[f_{j,k}]_{j,k=1,...,n}\,\Pf[g_{j,k}]_{j,k=1,...,N}, 
\end{eqnarray}
\begin{eqnarray}
N= \cases{n & for $n$ even,\\ n+1 & for $n$ odd.}
\end{eqnarray}
In the above expression, the kernels are 
 \begin{eqnarray}
 \label{f-def}
 f_{j,k}=f_{j}(\lambda_{k}) = \lambda_{k}^{j+\alpha-1}e^{-\lambda_{k}},
 \end{eqnarray}
 \begin{eqnarray}
 \label{gjk}
g_{j,k}=-g_{k,j}=g(\lambda_j,\lambda_k) =\frac{\lambda_k-\lambda_{j}}{\lambda_k+\lambda_j},
 \end{eqnarray}
and in addition, when $n$ is odd,
\begin{eqnarray}
g_{j,n+1}=-g_{n+1,j}=1-\delta_{j,n+1}.
 \end{eqnarray}
 The inverse of the normalisation factor (partition function) can be obtained using de Brujin's integration theorem~\cite{Bruijn1955} as
  \begin{eqnarray}
  \label{norm}
 C^{-1}= n! \,\Pf[H],
 \end{eqnarray}
 where $H$ is an $N$-dimensional matrix with elements,
  \begin{eqnarray}
  \label{Hjk}
  \fl
 H_{j,k}=\int_{0}^{\infty}d\lambda \int_{0}^{\infty} d\nu \, f_{j}(\lambda)f_{k}(\nu)g(\lambda,\nu) = \frac{k-j}{j+k+2\alpha} \Gamma(j+\alpha) \Gamma(k+\alpha),
\end{eqnarray}
for $1\leq j,k\leq n$, and additionally,
 \begin{eqnarray}
 \label{Hjnp1}
 \fl
 H_{j,n+1}=-H_{n+1,j}=(1-\delta_{j,n+1})\int_{0}^{\infty}d\lambda \, \, f_{j} (\lambda)=(1-\delta_{j,n+1})\Gamma(j+\alpha),
 \end{eqnarray}
 when $n$ odd.
As shown in the~\ref{AppPf}, the Pfaffian in~(\ref{norm}) can be evaluated to a yield a compact result for the normalization factor as~\cite{FK2016}
\begin{eqnarray}
\label{C-unres}
C= \frac{2^{n^2+2\alpha n}}{\pi^{n/2}}\prod_{j=1}^{n}\frac{\Gamma(j+\alpha+1/2)}{\Gamma(j+1)\Gamma(j+2\alpha+1)}.
\end{eqnarray}

Given the jpd of eigenvalues, one is interested in calculating the $r$-point correlation function, which is defined as
\begin{eqnarray}
R_{r}(\lambda_{1},..,\lambda_{r})=\frac{n!}{(n-r)!}\int_0^\infty d\lambda_{r+1}...\int_0^\infty d\lambda_{n} \mathcal{P}(\lambda_1,..,\lambda_{r},\lambda_{r+1},..,\lambda_{n}).
\end{eqnarray}
In the present case an exact result for $R_{r}$ can be obtained using the generalization of the de Brujin's theorem, as derived by Kieburg~\cite[Appendix A.1]{Kieburg2012}. The result is in terms of a Pfaffian of an $(N+2r)$-dimensional antisymmetric matrix:
\begin{eqnarray}
\label{Rr_U}
\fl
R_{r}(\lambda_1,..,\lambda_{r})=(-1)^{r(r-1)/2}n!\,C \, \, \Pf \left[\begin{array} {ccc} 
[0]_{j=1...r\atop k=1...r} &[0]_{j=1...r\atop k=1...r} & [F_{k j}]_{j=1...r\atop k=1...N}\\
 {[0]}_{j=1...r\atop k=1..r}  &[g_{j k}]_{j=1...r\atop k=1...r} &[G_{k j}]_{j=1...r\atop k=1...N}\\
 -[F_{jk}]_{j=1...N\atop k=1...r} &-[G_{jk}]_{j=1...N \atop k=1...r}& [H_{jk}]_{j=1...N\atop k=1...N}
\end{array}\right].
\end{eqnarray}
In the above expression, the $j,k$ indices in a matrix block $[\,\cdot\,]_{j=...\atop k=...}$ are the row and column indices, respectively. The kernels $G_{jk}$ and $F_{jk}$ appearing in the above Pfaffian are given by
\begin{eqnarray}
\label{Fjkev}
\fl
&F_{j,k}=F_j(\lambda_k)=f_j(\lambda_k)=\lambda_k^{j+\alpha-1}e^{-\lambda_k},\\
\fl
\label{Gjkev}
&G_{j,k}=G_j(\lambda_k)=\int_0^\infty d\nu \hspace{0.1cm}f_{j}(\nu)g(\nu,\lambda_{k})= \Gamma (j+\alpha) [2 \lambda_k \,e^{\lambda_k}\,\E_{j+\alpha}(\lambda_k)-1],
\end{eqnarray}
for $j=1,...,n; k=1,...,r$. Here E$_a(z)= \int_1^\infty dt\,e^{-zt}/t^a$ is the exponential integral function. Moreover, when $n$ is odd, we have
\begin{eqnarray}
\label{Fjkod}
&F_{n+1,k}=F_{n+1}(\lambda_k)=0,\\
\label{Gjkod}
&G_{n+1,k}=G_{n+1}(\lambda_k)=-1,
\end{eqnarray}
for $k=1,...,r$.
The $g_{jk}$ and $H_{jk}$ within the Pfaffian in~(\ref{Rr_U}) are as in (\ref{gjk}), (\ref{Hjk}) and (\ref{Hjnp1}). In reference~\cite{FK2016}, the $r$-point correlation function for the unrestricted Bures--Hall ensemble has been derived by exploiting its relationship with the Cauchy two-matrix ensemble. It involves the Pfaffian of a $2r\times 2r$ antisymmetric matrix with kernels involving integral over certain Meijer G-functions. While it appears difficult to demonstrate a direct equivalence of the Pfaffian result of~\cite{FK2016} with the above Pfaffian result, it can be numerically verified on a case-by-case basis that they are indeed equivalent.

\begin{figure}[!tbp]
\centering
\includegraphics[width = 5.5in ]{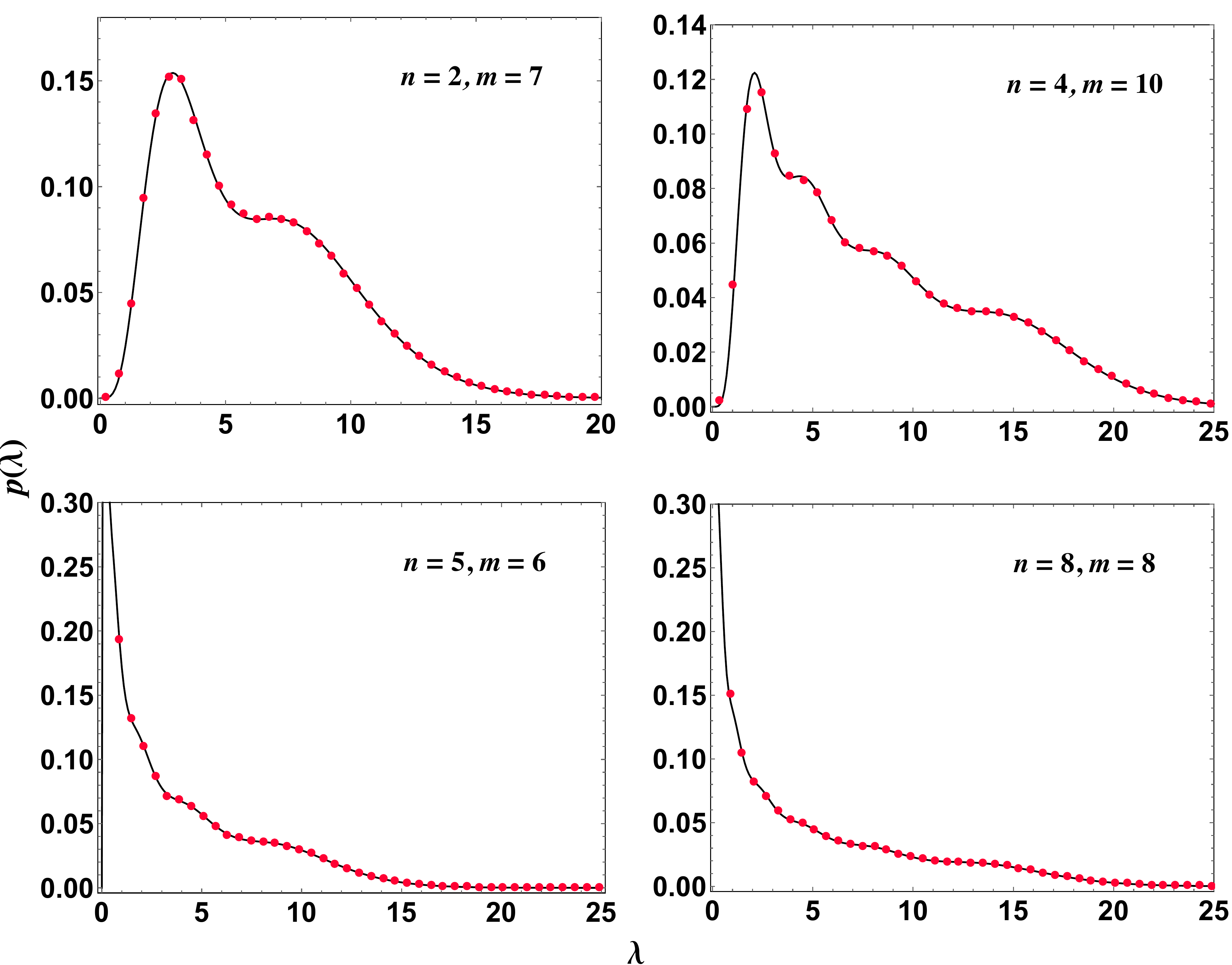}
\caption{Marginal density of eigenvalues for the unrestricted trace Bures--Hall ensemble for various values of $n$ and $m$. The solid lines are analytical predictions and the symbols denote the numerical result based on Dyson's log-gas approach.}
\label{FigUTdens}
\end{figure} 

The level density $R_1(\lambda)$ is of special interest since the first order marginal density $p(\lambda)=R_1(\lambda)/n$ reveals the behavior of a generic eigenvalue of the ensemble. Furthermore, it enables one to obtain the averages of observables which are linear statistic on the eigenvalues. We use the Pfaffian-expansion result given in~\cite[Corollary 2.4]{Okada2017} to obtain the following expression for the level density:
\begin{eqnarray}
\label{R1}
R_1(\lambda)=n! \,C\sum_{1\leq j < k \leq N} (-1)^{j+k}  \left[\Phi_{j,k}(\lambda)-\Phi_{k,j}(\lambda)\right] \,\Pf[H^{(j,k)}].
\end{eqnarray}
Here 
\begin{equation}
\Phi_{j,k}(\lambda)= F_{j}(\lambda)G_{k}(\lambda),
\end{equation}
and $H^{(j,k)}$ is the $(N-2)$-dimensional antisymmetric matrix obtained after removing the $j$th and $k$th rows and columns from $H$. For $n = 1$, $\Pf[H^{(1,2)}]$ should be taken as $1$. Moreover, as shown in~\ref{AppPf}, this Pfaffian can be evaluated in terms of a restricted product as
\begin{equation}
\label{restPf}
\Pf[H^{(j,k)}]=\prod_{r<s\atop r,s\ne j,k}\frac{s-r}{r+s+2\alpha}\cdot\prod_{l\ne j,k}\Gamma(l+\alpha).
\end{equation}
The notation $r,s\ne j,k$ in the product means that both $r$ and $s$ do not assume the values $j,k$.

In Fig.~\ref{FigUTdens} we show the plots of the marginal density $p(\lambda)$ for various values of $n, m$. The solid curves are based on the above analytical result, and the symbols have been obtained using the numerical simulation following Dyson's log-gas formalism~\cite{Mehta2004,Forrester2010}, as briefly described in the~\ref{AppGE}. We can see very good agreements between the analytical-expression based and numerical simulation based results.

\section{Fixed trace Bures--Hall ensemble} 
\label{SecFTBHE}

We now focus on the fixed trace Bures--Hall ensemble. The matrices constituting this ensemble are given by
\begin{equation}
\label{rho}
\rho=\frac{B}{\tr B},
\end{equation}
where $B$ are as in~(\ref{matrixB}). Here, we note that the choice of variance of the Gaussian matrix elements of $G$ constituting $B$ in~(\ref{PGG}) is immaterial and hence any $v^{2}>0$ leads to the same distribution. The above matrix model is equivalent to~(\ref{PBHM}) describing the reduced density matrices pertaining to the Bures--Hall measure. The related $m\times m$ dimensional and rank $n$ random matrix
 \begin{equation}
\label{rhop}
\rho'=\frac{B'}{\tr B'},
\end{equation}
where $B'$ is as in~(\ref{matrixBp}), share the $n$ eigenvalues of $\rho$ and additionally has $m-n$ zero eigenvalues.
The joint eigenvalue probability density of the eigenvalues ($\mu_j\in[0,1] ,j=1,...,n$) of $\rho$ in~(\ref{rho}), and hence of the $n\times n$ random density matrix appearing in~(\ref{PBHM}), is given by~\cite{Hall1998,ZS2001}
\begin{eqnarray}
\label{jpd-res}
\mathcal{P}^{(F)}(\mu_1,...,\mu_n) =C^{(F)} \frac{\Delta^2(\{\mu\})}{\Delta_{+}(\{\mu\})} \delta\Bigg(\sum_{i=1}^{n}\mu_{i}-1\Bigg)\prod_{j=1}^{n}\mu_{j}^{\alpha}.
\end{eqnarray}
The parameter $\alpha$, as discussed earlier, relates to the dimensions $m$ and $n$. For $n=m$, we have the standard Bures-Hall ensemble and  the distribution~(\ref{jpd-res}) of eigenvalues coincides with the one induced by the Bures metric over the space of random mixed states~\cite{ZPNC2011}. If we consider $ m = 1$, $n$ too must be $1$ as per our construction and this leads to $\alpha = -1/2$. The corresponding single eigenvalue then takes the value $1$ and physically corresponds to a pure state. In the limit of $m \to \infty$ with fixed $n$, $\alpha \to \infty$ and then, as discussed in the introduction, the $n$ eigenvalues approach the value of $1/n$. Physically, this signifies a maximally mixed state for $n>1$. 

It can be observed that, by introducing an auxiliary variable to replace the 1 inside the delta function, performing Laplace transform~\cite{Hoskins2009} and then applying some rescaling, we are led to the jpd given by~(\ref{jpd-unres}) for the unrestriced Bures--Hall ensemble. Consequently, the corresponding normalization factors are also related. As shown in the~\ref{AppCorr}, the normalization factor $C^{(F)}$ in the jpd~(\ref{jpd-res}) is given by
\begin{eqnarray}
\label{normFT}
\nonumber
C^{(F)}= \Gamma[n(n+2\alpha+1)/2]\,C\\
\nonumber
=\frac{2^{n(n+2\alpha)}\,\Gamma[n(n+2\alpha+1)/2]}{\pi^{n/2}}\prod_{j=1}^{n}\frac{\Gamma(j+\alpha+1/2)}{\Gamma(j+1)\Gamma(j+2\alpha+1)}\\
=\frac{2^{n(2m-n-1)}\,\Gamma[n(2m-n)/2]}{\pi^{n/2}}\prod_{j=1}^{n}\frac{\Gamma(j+m-n)}{\Gamma(j+1)\Gamma(j+2m-2n)}.
\end{eqnarray}
For the square case $m = n$, i.e. $\alpha=-1/2$, this reduces to the result conjectured by Slater in \cite{Slater1999} and later proved by Sommers and \.Zyczkowski in~\cite{SZ2003}, who also derived the above general result.
\begin{figure}[!h]
\centering
\includegraphics[width = 5.5in ]{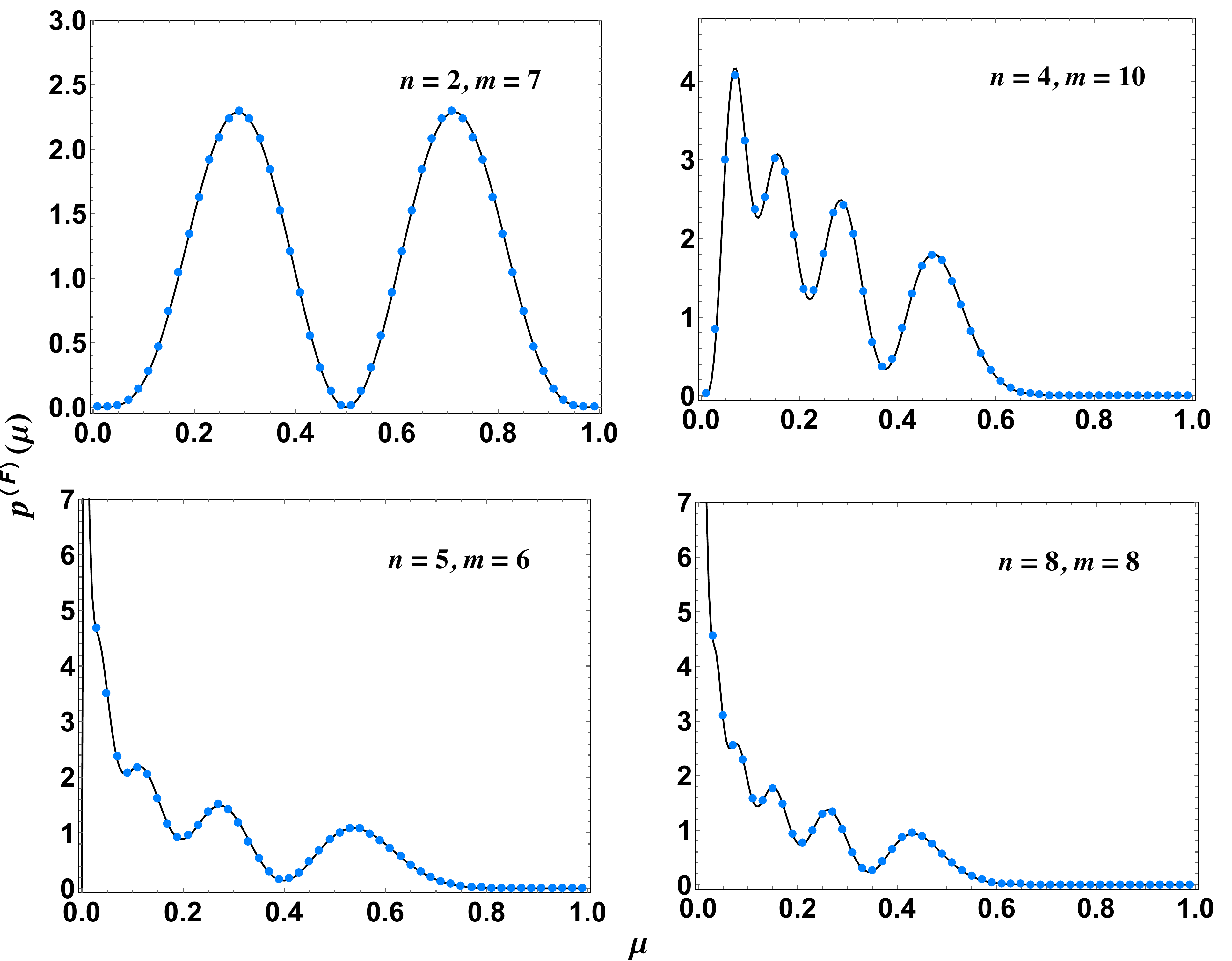}
\caption{Marginal density of eigenvalues for the fixed trace Bures--Hall ensemble for various $n$ and $m$ values. The solid lines are based on analytical results and the symbols are using numerical simulation based on Dyson's log-gas formalism.}
\label{FigFTdens}
\end{figure}

Similar to the jpd of eigenvalues, the $r$-level correlation function $R_{r}^{(F)}$ for the fixed trace ensemble can be related to that of the unrestricted trace ensemble $R_{r}$ by means of an inverse Laplace transform; see the~\ref{AppCorr}. We have the result
\begin{equation}
\label{CorrFuncRel}
\fl
R_{r}^{(F)}(\mu_1,...,\mu_r)= \Gamma[n(n+2\alpha+1)/2]\, \mathcal{L}^{-1}\left\{s^{r-n(n+2\alpha+1)/2} R_{r}(s\mu_1,...,s\mu_r)\right\}(t)\Big|_{t=1}.
\end{equation}
In particular, the level density $(r=1)$ for the fixed trace Bures--Hall ensemble can be obtained from unrestricted trace ensemble result using the relation
\begin{eqnarray}
\label{R1Fa}
R_{1}^{(F)}(\mu)=  \Gamma[n(n+2\alpha+1)/2]\,\mathcal{L}^{-1}\left\{s^{1-n(n+2\alpha+1)/2} R_{1}(s\mu)\right\}(t)\Big|_{t=1}.
\end{eqnarray}
Based on this Laplace inversion relationship and using (\ref{R1}), an exact closed form expression for the level density can be found. As shown in~\ref{AppR1F}, we obtain the non-vanishing result for $0\leq \mu\leq 1$ as
\begin{eqnarray}
\label{R1Fb}
R_{1}^{(F)}(\mu)=n!\,C^{(F)}\sum_{1\leq j < k \leq N} (-1)^{j+k} \left[\Psi_{j,k}(\mu)-\Psi_{k,j}(\mu)\right]\Pf[H^{(j,k)}].
\end{eqnarray}
In the above equation, $\Psi_{j,k}$ is given by
\begin{eqnarray}
\label{PsiEv}
\nonumber
\fl
&\Psi_{j,k}(\mu)= \Gamma(k+\alpha) \mu^{j+\alpha-1} \Big[\frac{2 \mu^{k+\alpha}\Gamma(1-k-\alpha)}{\Gamma(1-j-k-\alpha+\gamma)}\\
&~~~~~ - \frac{2\mu^{k+\alpha}\mathrm{B}_{\mu}(1-k-\alpha,\gamma-j)}{\Gamma(\gamma-j)}-\frac{(1-\mu)^{\gamma-j}}{\Gamma(\gamma-j+1)}\Big]; ~~~j,k=1,...,n,
\end{eqnarray}
and when $n$ is odd, additionally, we have
\begin{equation}
\label{PsiOd}
\Psi_{j,n+1}=-\frac{\mu^{j+\alpha-1}(1-\mu)^{\gamma-j}}{\Gamma(\gamma-j+1)},~\Psi_{n+1,k}(\mu)=0
;~~~j,k=1,...,n.
\end{equation}
Here, $\mathrm{B}_z(a,b)=\int_0^z du\, u^{a-1}(1-u)^{b-1}$ is the incomplete Beta function and, for compactness, we have defined
\begin{equation}
\label{gam}
\gamma = (n-1)(n+2\alpha+2)/2=(n-1) (2m-n+1)/2.
\end{equation}

In figure~\ref{FigFTdens} we show the plots of the marginal density $p^{(F)}(\mu)=R_1^{(F)}(\mu)/n$ for various combination of $n,m$ values. Again we find very good agreement between the analytical predictions (solid lines) and the numerical results (symbols) obtained using log-gas approach~\cite{Mehta2004,Forrester2010}, as discussed in~\ref{AppGE}.


\section{Entropy Measures}
\label{SecEM}

Given an ensemble of random density matrices, a natural question to ask is,"how far or close the associated states are to being pure or maximally mixed"? There are several entropy measures which can be used to answer this. We focus here on the Havrda-Charv\'at-Tsallis (HCT)~\cite{HC1967,Tsallis1988,BZ2006} entropy, which is defined as
\begin{eqnarray}
\mathcal{S}_{\omega}(\mu_1,...,\mu_n)= \frac{1}{\omega-1}\left(1-\sum_{i=1}^{n}\mu_{i}^{\omega}\right).
\end{eqnarray} 
Here $\omega\ne1$ is a positive real parameter. The value of $\mathcal{S}_{\omega}$ varies from 0 to $(1-n^{1-\omega})/(\omega-1)$. The former indicates a pure state, while the latter signifies a maximally mixed state.
In the limit $\omega \rightarrow 1$, the HCT entropy leads to the von Neumann entropy,
\begin{eqnarray}
\mathcal{S}_{1}(\mu_1,...,\mu_n)=-\sum_{i=1}^n \mu_{i} \ln\mu_{i}.
\end{eqnarray} 
For $\omega=2$, the HCT entropy yields the linear entropy,
\begin{eqnarray}
\mathcal{S}_{2}(\mu_1,...,\mu_n)= 1-\sum_{i=1}^n\mu_{i}^{2}.
\end{eqnarray}
The linear entropy $\mathcal{S}_{2}$ is related to the purity $\mathcal{S}_{P}$ as
\begin{equation}
\mathcal{S}_{P}(\mu_1,...,\mu_n) = 1-\mathcal{S}_{2}(\mu_1,...,\mu_n) = \sum_{i=1}^n\mu_{i}^{2}.
\end{equation}
 The HCT entropy is advantageous to use, as the ensemble averages are more easily done with $\sum_{i=1}^{n}\mu_{i}^{\omega}$ compared to the $\ln \left (\sum_{i=1}^{n}\mu_{i}^{\omega}\right)$ in the R\'enyi entropy~\cite{Renyi1961}.

\begin{table}[h!]
\renewcommand{\arraystretch}{1.5}
\caption{Exact values and numerical values (6 significant digits) for the average von Neumann entropy. }
\centering
{\small \begin{tabular}{|c|c|c|c|c|c| }
\hline
 \multirow{2}{*}{$n$} & \multirow{2}{*}{$m$} & \multicolumn{2}{|c|}{Bures--Hall}  & \multicolumn{2}{|c|}{Hilbert--Schmidt}   \\ 
 \cline{3-6}
 ~ &~ & Exact  & Numerical value & Exact & Numerical value  \\
\hline\hline
 2  & 2 & $2\ln 2-\frac{7}{6} $ & 0.2196277 & $\frac{1}{3}$ & 0.333333\\
 2  & 3 & $2\ln 2-\frac{59}{60}$ & 0.4029610 & $\frac{9}{20}$ & 0.450000\\
 2  & 4 & $2\ln 2-\frac{379}{420}$ & 0.4839134 & $\frac{107}{210}$ & 0.509524\\
 2  & 5 & $2\ln 2-\frac{2159}{2520}$ & 0.5295483 & $\frac{275}{504}$ & 0.545635\\
 2  & 6 & $2\ln 2-\frac{22937}{27720}$ & 0.5588413 & $\frac{15797}{27720}$ & 0.569877\\
 3  & 3 & $\frac{32}{63}$ & 0.5079365 & $\frac{1669}{2520}$ & 0.662302\\
 3  & 4 & $\frac{4448}{6435}$ & 0.6912199 & $\frac{21341}{27720}$ & 0.769877\\
 3  & 5 & $\frac{1272512}{1616615}$ & 0.7871460 & $\frac{300863}{360360}$ & 0.834896\\
 3  & 6 & $\frac{386215616}{456326325}$ & 0.8463584 & $\frac{239175}{272272}$ & 0.878441\\
 4  & 4 & $2\ln 2-\frac{533}{840}$ & 0.7517706 & $\frac{664789}{720720}$ & 0.922396\\
 4  & 5 & $2\ln 2-\frac{13067}{27720}$ & 0.9149019 & $\frac{15743083}{15519504}$ & 1.014406\\
 4  & 6 & $2\ln 2-\frac{270769}{720720}$ & 1.010602 & $\frac{1920308783}{1784742960}$ & 1.075958\\
 5  & 5 & $\frac{177377888}{185910725}$ & 0.9541025 & $\frac{10107221087}{8923714800}$ & 1.132625\\
 5  & 6 & $\frac{4952992040384}{4512611027925}$ & 1.097589 & $\frac{2822050213687}{2329089562800}$ & 1.211654\\
 6  & 6 & $2\ln 2-\frac{3201673}{12252240}$ & 1.124981& $\frac{17169484377589}{13127595717600}$ & 1.307893\\
\hline
\end{tabular}}
\label{tableVN}
\qquad
\end{table}

\begin{table}[h!]
\renewcommand{\arraystretch}{1.5}
\caption{Exact values and numerical values (6 significant digits) for the average Purity. }
\centering
{\small \begin{tabular}{|c|c|c|c|c|c| }
\hline
 \multirow{2}{*}{$n$} & \multirow{2}{*}{$m$} & \multicolumn{2}{|c|}{Bures--Hall}  & \multicolumn{2}{|c|}{Hilbert--Schmidt}   \\ 
 \cline{3-6}
 ~ &~ & Exact  & Numerical value & Exact & Numerical value \\
\hline\hline
 2  & 2 & $\frac{7}{8} $ & 0.875000 & $\frac{4}{5}$ & 0.800000\\
 2  & 3 & $\frac{3}{4}$ & 0.750000 & $\frac{5}{7}$ & 0.714286\\
 2  & 4 & $\frac{11}{16}$ & 0.687500 & $\frac{2}{3}$ & 0.666667\\
 2  & 5 & $\frac{13}{20}$ & 0.650000 & $\frac{7}{11}$ & 0.636364\\
 2  & 6 & $\frac{5}{8}$ & 0.625000 & $\frac{8}{13}$ & 0.615385\\
 3  & 3 & $\frac{23}{33}$ &0.696970 & $\frac{3}{5}$ & 0.600000\\
 3  & 4 & $\frac{10}{17}$ & 0.588235 & $\frac{7}{13}$ & 0.538462\\
 3  & 5 & $\frac{61}{115}$ & 0.530435 & $\frac{1}{2}$ & 0.500000\\
 3  & 6 & $\frac{43}{87}$ & 0.494253 & $\frac{9}{19}$ & 0.473684\\
 4  & 4 & $\frac{9}{16}$ & 0.562500 & $\frac{8}{17}$ & 0.470588\\
 4  & 5 & $\frac{25}{52}$ & 0.480769 & $\frac{3}{7}$ & 0.428571\\
 4  & 6 & $\frac{59}{136}$ & 0.433824 & $\frac{2}{5}$ & 0.400000\\
 5  & 5 & $\frac{7}{15}$ & 0.466667 & $\frac{5}{13}$ & 0.384615\\
 5  & 6 & $\frac{15}{37}$ & 0.405405 & $\frac{11}{31}$ &0.354839\\
 6  & 6 & $\frac{181}{456}$ & 0.396930& $\frac{12}{37}$ & 0.324324\\
\hline
\end{tabular}}
\label{tablePur}
\qquad
\end{table}

The calculation of average entropy can be performed in two ways. In the first approach, the result for the average entropy associated with the Bures--Hall ensemble can be found by directly integrating with the fixed trace ensemble jpd~(\ref{jpd-res}),
\begin{eqnarray}
\label{avHCT}
\langle \mathcal{S}_{\omega}\rangle_{\mathrm{BH}}= \int_{0}^{1}\cdots\int_{0}^{1} \mathcal{S}_{\omega}(\mu_1,...,\mu_n)\, \, \mathcal{P}^{(F)}(\mu_1,...,\mu_n) \,d\mu_{1}...d\mu_{n}.
\end{eqnarray}
Now, HCT entropy being a linear statistic, the symmetry of the eigenvalues in the jpd allows us to reduce the above average involving $n$ integrals to an average involving a single integral on the  level density $R_{1}^{(F)}(\mu)$. We obtain
\begin{eqnarray}
\label{avHCTa}
\langle \mathcal{S}_{\omega}\rangle_{\mathrm{BH}} = \frac{1}{\omega-1}-\frac{1}{\omega-1}\int_{0}^{1} \mu^{\omega}\, \, R^{(F)}_{1}(\mu) d\mu.
\end{eqnarray}
The second approach relies on mapping the average entropy calculation as an average over the level density of the unrestricted trace ensemble. This has been done in the~\ref{AppHCTun}, and yields the result
 \begin{eqnarray}
 \label{avHCTb}
\langle \mathcal{S}_{\omega}\rangle_{\mathrm{BH}}&=\frac{1}{\omega-1}-\frac{ \Gamma(\alpha+\gamma+1)}{(\omega-1)\,\Gamma(\alpha+\gamma+\omega+1)}\int_{0}^{\infty}\lambda^{\omega}\,R_{1}(\lambda)\, d\lambda .
\end{eqnarray} 
As demonstrated in the~\ref{AppHCT}, after some simplification and rearrangement, both approaches lead to the same expression, which is given by
\begin{eqnarray}
\label{avSomg}
\nonumber
\langle \mathcal{S}_{\omega}\rangle_{\mathrm{BH}}= \frac{1}{\omega-1}&-&\frac{n!\, C^{(F)}}{(\omega-1)\,\Gamma(\alpha+\gamma+\omega+1)}\\
&&\times \sum_{1\leq j<k\leq N} (-1)^{j+k} (\eta_{j,k}-\eta_{k,j} ) \Pf \left[H^{(j,k)}\right],
\end{eqnarray}
with
\begin{eqnarray}
\label{etaev}
\eta_{j,k}=\left(\frac{j-k+\omega}{j+k+2\alpha+\omega}\right)\Gamma(j+\alpha+\omega)\Gamma(k+\alpha);~~j,k=1,...,n.
\end{eqnarray}
Moreover, when $n$ is odd, we have
\begin{equation}
\label{etaod}
\eta_{j,n+1} =-\Gamma(j+\alpha+\omega),\, \eta_{n+1,k} = 0; ~~j,k=1,...,n.
\end{equation}

For $\omega=2$,~(\ref{avSomg}) gives the average linear entropy $\langle \mathcal{S}_{2} \rangle_{\mathrm{BH}}$, and also the average purity via the relation $\langle \mathcal{S}_{P} \rangle_{\mathrm{BH}}=1-\langle \mathcal{S}_{2} \rangle_{\mathrm{BH}}$. The average von Neumann entropy can be obtained by carefully taking the limit $\omega \rightarrow 1$, and after some rearrangement of terms, as
\begin{eqnarray}
\label{avS1}
\fl
\langle \mathcal{S}_{1}\rangle_{\mathrm{BH}} &= \psi(\alpha+\gamma+2)
-\frac{n!\, C^{(F)}}{\Gamma(\alpha+\gamma+2)}\sum_{1\leq j<k\leq N} (-1)^{j+k} (\xi_{j,k}-\xi_{k,j})\Pf\left[H^{(j,k)}\right],
\end{eqnarray}
where
\begin{eqnarray}
\label{xjkev}
\fl
\xi_{j,k}= \frac{(j-k+1)}{(j+k+2\alpha+1)}\Gamma (j+\alpha +1) \Gamma (k+\alpha )\psi (j+\alpha +1);~~j,k=1,...,n,
\end{eqnarray}
along with
\begin{equation}
\label{xjnp1}
\xi_{j,n+1} =-\Gamma (j+\alpha +1)\psi (j+\alpha +1),~\xi_{n+1,k} = 0; ~~~j,k=1,...,n,
\end{equation}
when $n$ is odd. Here $\psi(y)$ is the digamma function with the integral representation $\psi(y) = [1/\Gamma(y)]\int_{0}^{\infty} e^{-r} r^{y-1} \ln r \,dr$.

\begin{figure}[!ht]
\centering
\includegraphics[width = 6in ]{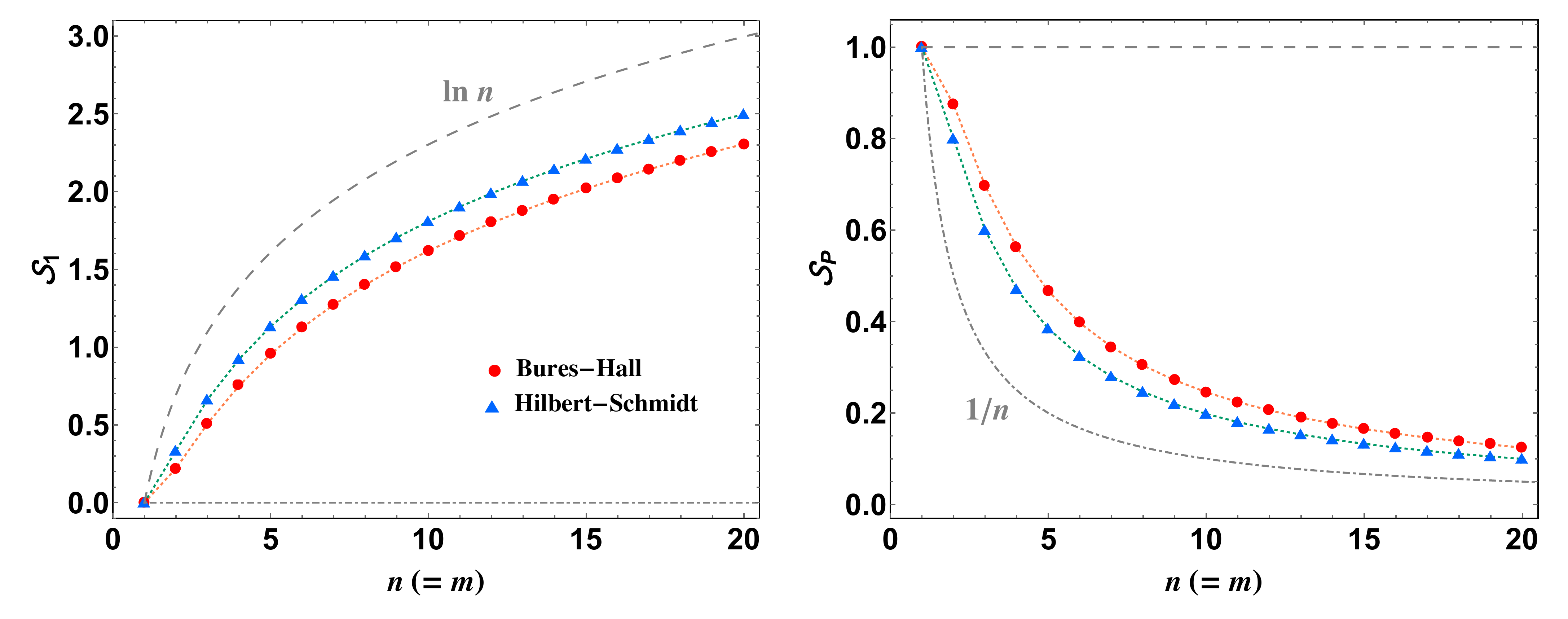}
\caption{Average von Neumann entropy and average purity for the square case ($m=n$) with varying $n$ values. Results for both Hilbert--Schmidt ensemble and Bures--Hall ensemble are shown. The von Neumann entropy varies from 0 for a pure state to $\ln n$ for a maximally mixed state. The purity varies from $1/n$ for a maximally mixed state to 1 for a pure state.}
\label{FigAvEntSq}
\end{figure}
\begin{figure}[!h]
\centering
\includegraphics[width = 6in ]{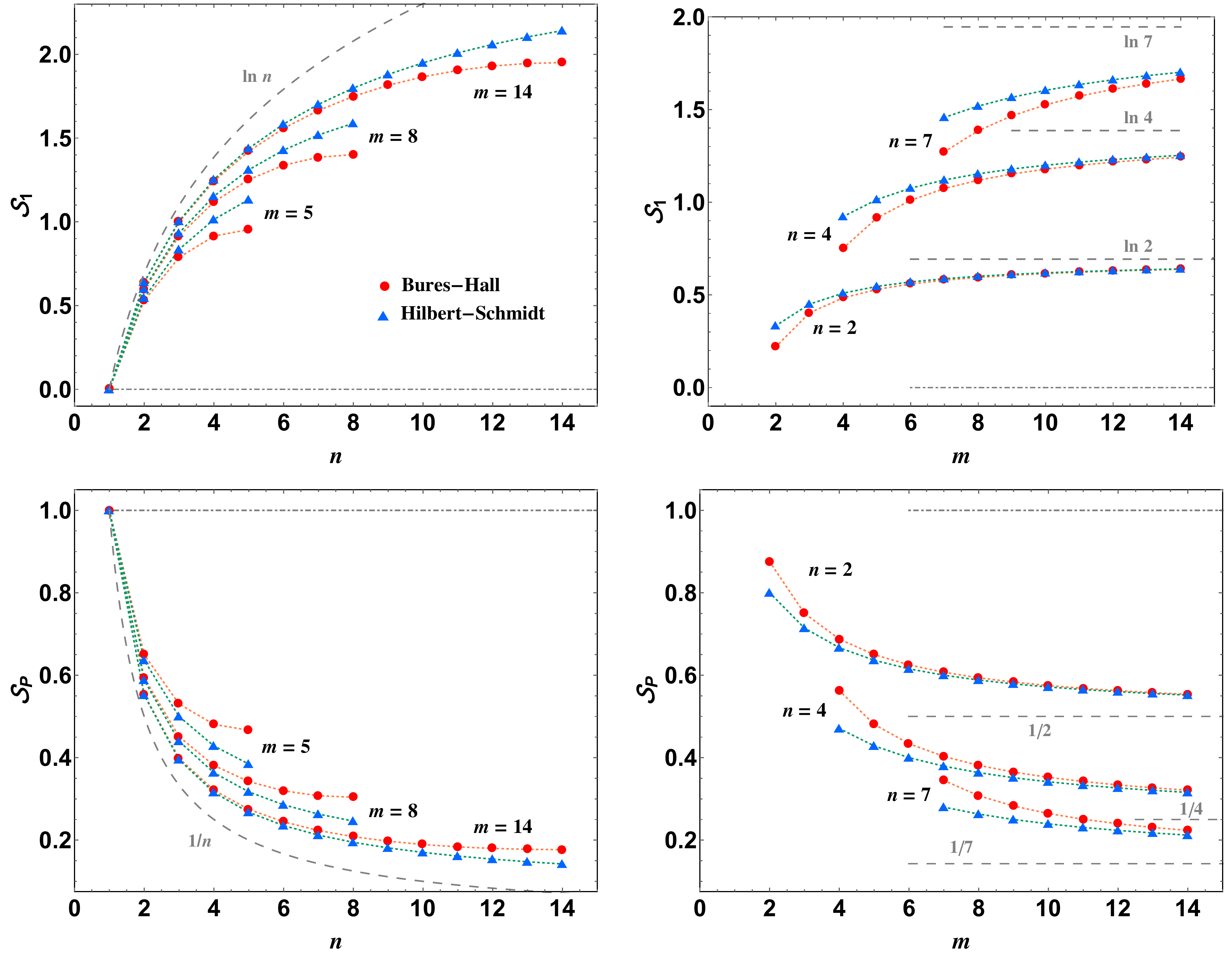}
\caption{Average von Neumann entropy and average purity for various combination of $n,m$ values. Results for both Hilbert--Schmidt and Bures--Hall ensembles are depicted.}
\label{FigAvEnt}
\end{figure}

In Tables~\ref{tableVN} and~\ref{tablePur} we compile the evaluations for average von Neumann entropy $\langle\mathcal{S}_1\rangle_{\mathrm{BH}}$, and the average purity $\langle \mathcal{S}_P\rangle_{\mathrm{BH}}=1-\langle\mathcal{S}_2\rangle_{\mathrm{BH}}$. For comparison, we also show the results for Hilbert--Schmidt ensemble, for which the average von Neumann entropy and average purity are given by~\cite{Page1993,Sen1996,SZ2004,Giraud2007}
\begin{equation}
\label{AvS1HS}
\langle\mathcal{S}_1\rangle_{\mathrm{HS}}=\sum_{j=m+1}^{mn}\frac{1}{j}-\frac{n-1}{2m}=\psi(m n+1)-\psi(m+1)-\frac{n-1}{2m},
\end{equation}
\begin{equation}
\label{AvSpHS}
\langle\mathcal{S}_P\rangle_{\mathrm{HS}}=\frac{m+n}{mn+1}.
\end{equation}
We should emphasise here that it is reasonable to compare the above entropic quantities for the two ensembles. This is because in either case $n$ signifies the dimension of the density matrix and $m$ refers to the dimension of the auxiliary subsystem (environment), as discussed in the introduction for partial tracing approach to induced measure over random density matrices.
 
In figures~\ref{FigAvEntSq} and~\ref{FigAvEnt} we show the average entropy results for $\mathcal{S}_1$ (von Neumann entropy) and $\mathcal{S}_P=1-\mathcal{S}_2$ (purity) for both Bures--Hall and Hilbert--Schmidt ensembles. The former figure depicts behavior of the entropies for the square ($m=n$) case which corresponds to the standard Hilbert--Schmidt and Bures--Hall ensembles and is of a special significance~\cite{BZ2006,ZS2001}. The latter deals with the general case and displays behavior of the entropies for $n$ fixed, $m$ varying, and $m$ fixed, $n$ varying scenarios. We find that for any $n\ge 2$, on average, the Bures--Hall measure is concentrated more towards states of higher purity than the Hilbert--Schmidt measure. This was shown in~\cite{SZ2004} for the square case and now we see that it holds even for the rectangular case. Additionally, it can be seen that for fixed $m$, the difference in averages increases as $n$ is increased towards $m$. On the other hand, for fixed $n$, if the $m$ value is increased, then the difference approaches zero.

On examining the exact evaluations of the average von Neumann entropy and the average purity (or, equivalently, the average linear entropy) for the Bures--Hall ensemble, we come up with the conjecture that, similar to those of the Hilbert--Schmidt ensemble, they are given by very simple formulae:
\begin{eqnarray}
\label{AvS1BH}
\langle\mathcal{S}_1\rangle_{\mathrm{BH}}&=\psi(m n-n^2/2+1)-\psi(m+1/2)\\
&=\cases{
\sum\limits_{j=1}^{mn-n^2/2}\frac{1}{j}-\sum\limits_{j=1}^m\frac{1}{j-1/2}+2\ln 2, & $n$ even,\\
\sum_{j=m+1}^{mn-(n^2-1)/2}\frac{1}{j-1/2}, & $n$ odd.
}
\end{eqnarray}
\begin{equation}
\label{AvSpBH}
\langle\mathcal{S}_P\rangle_{\mathrm{BH}}=\frac{2 m (2 m+n)-(n^2-1)}{2 m \left(2 m n-n^2+2\right)}, 
\end{equation}
or equivalently,
\begin{equation}
\langle\mathcal{S}_2\rangle_{\mathrm{BH}}=\frac{(n-1)(2m-1)(2m-n-1)}{2 m \left(2 m n-n^2+2\right)}.
\end{equation}
Comparing these conjectural expressions with~(\ref{avS1}) and (\ref{avSomg}), we find that one way to prove the above conjectures is to demonstrate the following equalities:
\begin{eqnarray}
\sum_{1\leq j<k\leq N} (-1)^{j+k} (\xi_{j,k}-\xi_{k,j})\Pf\left[H^{(j,k)}\right]= \frac{(mn-n^2/2)}{n!\, C}\psi(m+1/2),
\end{eqnarray}
\begin{eqnarray}
\fl
 \sum_{1\leq j<k\leq N} (-1)^{j+k} (\eta_{j,k}-\eta_{k,j} ) \Pf \left[H^{(j,k)}\right]=\frac{n(2m-n)}{n!\, C}\cdot\frac{2 m (2 m+n)-(n^2-1)}{8 m},
\end{eqnarray}
where $\xi_{j,k}$ is given by (\ref{xjkev}), (\ref{xjnp1}), and $\eta_{j,k}$ is given by (\ref{etaev}), (\ref{etaod}) with $\omega$ set to 2.

It readily follows from~(\ref{AvS1HS}), (\ref{AvSpHS}), (\ref{AvS1BH}) and (\ref{AvSpBH}) that in the limit $m\to \infty$ the average von-Neumann entropy and average purity for both Hilbert--Schmidt and Bures--Hall ensembles assume the values $\ln n$ and $1/n$, respectively. These values are expected in view of the discussion in the introduction concerning the $m\to\infty$ limit with $n$ fixed. Moreover, we find that
\begin{equation}
\langle\mathcal{S}_P\rangle_{\mathrm{BH}} - \langle\mathcal{S}_P\rangle_{\mathrm{HS}}=\frac{(mn-1)(n^2-1)}{2m(mn+1)(2mn-n^2+2)}\ge0,
\end{equation}
where the equality holds for $n=1$. Therefore, this is in conformity with the conclusion that on average the Bures--Hall measure is concentrated more towards the states of higher purity compared to the Hilbert--Schmidt measure. The other observations found in Fig.~\ref{FigAvEnt} for the difference of average purities are also consistent with this expression.

\section{Summary and Outlook}
\label{SecSum}

In this work we investigated statistical properties of random matrix ensemble distributed according to the Bures measure. We started our analysis with the unrestricted trace Bures--Hall ensemble and obtained a new Pfaffian based representation for the correlation function of arbitrary order. This was then used to obtain a closed-form exact result for the level density of the fixed trace Bures--Hall ensemble. Based on this, we computed the average HCT entropy and also obtained the average von-Neumann entropy, average linear entropy and average purity by considering appropriate limits. The exact evaluations of these average entropies enabled us to propose very simple conjectural expressions for the average von-Neumann entropy and average linear entropy or purity. We found that the Bures measure is concentrated more towards states of higher purity than the Hilbert--Schmidt measure even for the rectangular case.

The simple conjectural expressions suggest that there is some additional structure in the eigenvalue statistics of Bures-Hall ensemble that needs to be unveiled. Moreover, it would be of interest to obtain higher order statistics for the entropies, i.e, higher moments and cumulants. This would help provide a better understanding concerning the statistics of these entropies when dealing with Bures measure.

\section{Acknowledgement}
AS acknowledges DST-INSPIRE for providing the research fellowship [IF170612]. SK is grateful to Prof P J Forrester for fruitful correspondences.


\appendix

\section{\label{AppPf} Evaluation of Pfaffians}
We need to evaluate $\Pf[H]$, where $H$ is an antisymmetric matrix defined by~(\ref{Hjk}) and~(\ref{Hjnp1}). First of all, we observe that factors $\Gamma(i+\alpha)$, $i=1,...,n,$ can be pulled out of the Pfaffian from the $i$th row as well as the $i$th column, thereby giving rise to an overall factor of $\prod_{i=1}^n \Gamma(i+\alpha)$. Therefore, we obtain
\begin{equation}
\fl
\Pf[H]=\prod_{i=1}^n \Gamma(i+\alpha)\cdot \cases{
\Pf\left[(k-j)/(k+j+2\alpha)\right]_{j,k=1,...,n}, & $n$  even,\\
\Pf\left[\begin{array}{cc}
\left[(k-j)/(k+j+2\alpha)\right]_{j,k=1,...,n} & \left[1\right]_{j=1,...,n} \\
\left[-1\right]_{k=1,...,n} & 0
\end{array}\right], & $n$  odd,
}
\end{equation}
Comparing this with the Schur's Pfaffian identity, equation~(\ref{SchurPf}), we find that by setting $x_i=i+\alpha$, $i=1,...,n$, we can reduce the above Pfaffian to the following product:
\begin{equation}
\fl
\Pf[H]=\prod_{i=1}^n \Gamma(i+\alpha)\cdot\prod_{1\le j<k\le n} \frac{k-j}{k+j+2\alpha}=\prod_{i=1}^n \Gamma(i+\alpha)\cdot\prod_{k=2}^{n}\prod_{j=1}^{k-1} \frac{k-j}{k+j+2\alpha}.
\end{equation}
Now, $\prod_{j=1}^{k-1} (k-j)=(k-1)(k-2)\cdots(2)(1)=\Gamma(k)$, and $\prod_{j=1}^{k-1} (k+j+2\alpha)=(k+2\alpha+1)(k+2\alpha+2)\cdots(2k+2\alpha-1)=\Gamma(2k+2\alpha)/\Gamma(k+2\alpha+1)$.
Therefore, we have
\begin{equation}
\fl
\Pf[H]=\prod_{i=1}^n \Gamma(i+\alpha)\cdot\prod_{k=2}^{n}\frac{\Gamma(k)\Gamma(k+2\alpha+1)}{\Gamma(2k+2\alpha)}=\prod_{i=1}^n \Gamma(i+\alpha)\cdot\prod_{k=1}^{n}\frac{\Gamma(k)\Gamma(k+2\alpha+1)}{\Gamma(2k+2\alpha)},
\end{equation}
where the last step followed because the $k=1$ term in the second product is 1. We now use the identity $\Gamma(2k+2\alpha)=2^{2k+2\alpha-1}\Gamma(k+\alpha+1/2)\Gamma(k+\alpha)/\pi^{1/2}$. This yields
\begin{equation}
\fl
\Pf[H]=\frac{\pi^{n/2}}{2^{n^2+2\alpha n}}\prod_{k=1}^{n}\frac{\Gamma(k)\Gamma(k+2\alpha+1)}{\Gamma(k+\alpha+1/2)}=\frac{\pi^{n/2}}{2^{n^2+2\alpha n}\,n!}\prod_{k=1}^{n}\frac{\Gamma(k+1)\Gamma(k+2\alpha+1)}{\Gamma(k+\alpha+1/2)},
\end{equation}
which is the required result.

For the evaluation of $\Pf[H^{(j,k)}]$ as a restricted product given in~(\ref{restPf}), first of all, we pull out the $\Gamma(i+\alpha)$, $i=1,...,n; i\ne j,k$ from the Pfaffian. This gives a factor $\prod_{i=1; i\ne j,k}^n \Gamma(i+\alpha)$. Now, we use Schur's Pfaffian identity, equation~(\ref{SchurPf}), with $\{x_1,...,x_{n-2}\}=\{s+\alpha \,\big|\, s=1,...,n$; $s\neq j,k$\}. This gives the result~(\ref{restPf}).


\section{\label{AppGE} Generating eigenvalues from the Bures--Hall ensemble}

Dyson showed that the jpd of eigenvalues of classical random matrix ensembles coincide with the Gibbs-Boltzmann factor associated with a log-gas system at three special values of the inverse temperature $\beta = 1, 2$ and 4, with the Boltzmann constant set to 1~\cite{Mehta2004,Forrester2010}. The `log' has to do with the fact that the gas particles interact via Coulombic interaction, which is logarithmic in two-dimensional space. We can use the same idea here and, for the unrestricted ensemble case, interpret the jpd in~(\ref{jpd-unres}) as a Gibbs-Boltzmann weight:
\begin{equation}
\mathcal{P}(\lambda_1,\dots,\lambda_n)=C e^{-\beta W},
\end{equation}
with $\beta=2$. This gives the energy function as
\begin{eqnarray}
\nonumber
&W=-\frac{1}{2}\ln\left(\frac{\Delta^2(\{\lambda\})}{\Delta_+(\{\lambda\})}\prod_{i=1}^n \lambda_i^\alpha e^{-\lambda_i}\right)\\
&=\frac{1}{2}\left(-2\sum_{j<k} \ln|\lambda_j-\lambda_k|+\sum_{j<k} \ln|\lambda_j+\lambda_k|-\alpha\sum_{i=1}^n \ln\lambda_i+\sum_{i=1}^n \lambda_i\right).
\end{eqnarray}
In this case, along with the two-dimensional Coulombic interaction, there is another two-body interaction between the particles given by $\ln|\lambda_j+\lambda_k|$. Moreover, the particles are constrained to move on the positive real axis, and also feel the one-body potential $-\alpha \ln\lambda_i+\lambda_i$. The above energy function can be implemented in the standard Metropolis-Hastings algorithm based Monte Carlo simulation to generate the stationary configurations, which yield the jpd. The generic one-eigenvalue density $p(\lambda)$ can be obtained by collecting the values of all $\lambda_i$ and then plotting the histogram. We note that numerically we have to put a ``large" cut-off for the domain of eigenvalues (position of the particles) as we cannot assign the full positive real domain. It should be large enough so that the eigenvalues density resulting from the simulation becomes negligible beyond this cut-off.

For the eigenvalues in the fixed trace case, we can use the eigenvalues $\lambda_i$ generated in the above simulation, and obtain $\mu_i=\lambda_i/(\sum_{j=1}^n \lambda_j)$. Another option is to implement the simulation with the above energy function but restrict positions of the charges ($\mu_i$) in the domain $[0,1]$, along with the constraint that $\sum_{i=1}^n \mu_i=1$. In this case, for the Metropolis-Hastings algorithm, we perturb positions of two charges simultaneously by amounts $\delta \mu$ and $-\delta \mu$, so that the fixed trace constraint remains imposed throughout the simulation, if one starts with such an initial configuration~\cite{NMV2011}. Additionally, if position of any of the two charges fall outside $[0,1]$ as a result of perturbation, that move is rejected. In the present work, we have used the first approach as it generates the eigenvalue densities for both unrestricted and restricted trace ensembles at once.

Yet another way to obtain the eigenvalues is to use the random matrix itself using~(\ref{matrixB}) and~(\ref{rho}) for the unrestricted trace and fixed trace cases, respectively, and then diagonalize it. For the square case ($n=m$), it is comparatively easier to do so since one requires generating complex Ginibre (Gaussian) matrices, and Haar-distributed unitary matrices. For the rectangular case ($n<m$), the unitary matrices have to be generated from the measure $|\mathrm{det}(\mathds{1}_n + U)|^{2(m-n)}d\mu(U_n)$, and this requires some additional work. One way to do this is to implement Monte Carlo simulation by performing random walk in the space of unitary matrices and use Metropolis-Hastings algorithm with the statistical weight $|\mathrm{det}(\mathds{1}_n + U)|^{2(m-n)}$, i.e. use the energy function as $-2(m-n)\ln|\mathrm{det}(\mathds{1}_n + U)|$. The perturbation moves in the unitary matrices can be implemented using $U\to U\exp(i\delta M)$, where $M$ is an $n\times n$-dimensional Hermitian random matrix from the Gaussian Unitary Ensemble (GUE)~\cite{Mehta2004,Forrester2010}, and $\delta$ is a small scalar. Once the stationarity is achieved, the unitary matrices would be generated from the measure $|\mathrm{det}(\mathds{1}_n+ U)|^{2(m-n)}d\mu(U_n)$ and hence can be used to construct $(\ref{matrixB})$ and~$(\ref{rho})$, which can then be diagonalized to obtain the eigenvalues. It should be noted that the eigenvalues for $\rho$ can be obtained easily using those of $B$ and therefore, in practice, one needs to diagonalize $B$ only.

Finally, in view of the invariant nature of the measure $|\mathrm{det}(\mathds{1}_n + U)|^{2(m-n)}d\mu(U_n)$, we may first generate the corresponding eigenangles $\{\vartheta_j\}$ using the log-gas approach and then obtain matrices $U$ by conjugating $\mathrm{diag}(\vartheta_1,\dots,\vartheta_n)$ with unitary matrices from the Haar measure.

\section{Relationship between level densities of fixed trace and unrestricted trace ensembles}
\label{AppCorr}

Establishing a relationship between the fixed trace and unrestricted trace variants of the Bures--Hall ensemble is based on implementing a Laplace transform which has been used in earlier works also. We use the same idea here and introduce an auxiliary variable $t$ to replace the 1 for the fixed trace condition in the expression for $r$-level density function for the fixed trace Bures--Hall ensemble:
\begin{eqnarray}
\nonumber
R_{r}^{(F)} (\mu_1,...,\mu_{r};t)&=\frac{n!}{(n-r)!}\int_{0}^{\infty}d\mu_{r+1}...\int_{0}^{\infty}d\mu_{n} C^{(F)}\\
&\times\delta\left(\sum_{i=1}^{n}\mu_{i}-t\right) \frac{\Delta^{2}(\{\mu\})}{\Delta_{+}(\{\mu\})}\prod_{j=1}^{n} \mu^{\alpha}_{j}.
\end{eqnarray}
It should be noted that we have extended the integration domains from $[0,1]$ to $(0,\infty)$ and this keeps the result of the multidimensional integral unchanged due to the delta-function constraint. We now apply the Laplace transform ($t\to s$) and obtain
\begin{eqnarray}
\fl
\widetilde{R}_{r}^{(F)}(\mu_1...\mu_{r},s)&=\frac{n!}{(n-r)!}\int_{0}^{\infty}d\mu_{r+1}....\int_{0}^{\infty}d\mu_{n} \, C^{(F)}\frac{\Delta^{2}(\{\mu\})}{\Delta_{+}(\{\mu\})}\prod_{j=1}^{n} \mu^{\alpha}_{j}e^{-s \mu_{j}}.
\end{eqnarray}
After some rearrangement, the right hand side of the above equation can be expressed in terms of the level density of the unrestricted trace ensemble as
\begin{eqnarray}
\widetilde{R}_{r}^{(F)}(\mu_1,...\mu_{r}, s)= \frac{C^{(F)}}{C}\frac{1}{s^{n(n+2\alpha+1)/2-r}} R_{r}(s\mu_1,...,s\mu_r),
\end{eqnarray}
so that the application of inverse-Laplace transform and substitution of $t=1$ yields
\begin{equation}
\label{RrF}
R_{r}^{(F)}(\mu_1,...\mu_{r})= \frac{C^{(F)}}{C}\mathcal{L}^{-1}\left\{\frac{1}{s^{n(n+2\alpha+1)/2-r}} R_{r}(s\mu_1,...,s\mu_r)\right\}(t)\Bigg|_{t=1}.
\end{equation}
To obtain the ratio $C^{(F)}/C$ we can set $r=0$ in the above expression, which for the correlation function means integrating the jpd over all eigenvalues and multiplying with $n!/(n-0)!$ and therefore gives the result $R_0=1$. As a consequence, we have
\begin{equation}
1=\frac{C^{(F)}}{C}\mathcal{L}^{-1}\left\{\frac{1}{s^{n(n+2\alpha+1)/2}}\right\}(t)\Bigg|_{t=1},
\end{equation}
which upon performing the inverse Laplace transform yields
\begin{equation}
\label{normratio}
\frac{C^{(F)}}{C}=\Gamma[n(n+2\alpha+1)/2]=\Gamma(\alpha+\gamma+1)=\Gamma(mn-n^2/2),
\end{equation}
where $\gamma=(n-1)(n+2\alpha+2)/2$ as in~(\ref{gam}). Plugging this back in~(\ref{RrF}) gives~(\ref{CorrFuncRel}). 

\section{Level density for the fixed trace ensemble using inverse Laplace transform}
\label{AppR1F} 

We use the expression for $R_1(\lambda)$ from~(\ref{R1}) into~(\ref{R1Fa}), along with the relationship~(\ref{normFT}) between the normalization factors. The inverse Laplace transform acts on the relevant parts of the full expression:
\begin{eqnarray}
\label{R1Fapp}
\nonumber
R_{1}^{(F)}(\mu)=n! C^{(F)}&&\sum_{j<k}(-1)^{j+k}[\mathcal{L}^{-1}\{s^{1-n(n+2\alpha+1)/2}\Phi_{j,k}(s\mu)\}(t)\\
&&-\mathcal{L}^{-1}\{s^{1-n(n+2\alpha+1)/2}\Phi_{k,j}(s\mu)\}(t)]\big|_{t=1}.
\end{eqnarray}
Therefore, we essentially seek the inverse Laplace transform of $s^{1-n(n+2\alpha+1)/2}\Phi_{j,k}(s\mu)=s^{1-n(n+2\alpha+1)/2}F_j(s\mu)G_k(s\mu)$, where $F_j(\lambda)$ and $G_k(\lambda)$ are as in equations~(\ref{Fjkev})--(\ref{Gjkod}). This requires the following results:
\begin{eqnarray}
\mathcal{L}^{-1}\{s^{-a}e^{-s\mu}\}(t)\big|_{t=1}=\frac{(1-\mu)^{a-1}\Theta(1-\mu)}{\Gamma(a)},\\
\mathcal{L}^{-1}\{s^{-a}E_b(s\mu)\}(t)\big|_{t=1}=\frac{\mu^{b-1}}{\Gamma(a)}[\mathrm{B}(1-b,a)-\mathrm{B}_\mu(1-b,a)]\Theta(1-\mu).
\end{eqnarray}
where $\Theta(u), \mathrm{B}(u,v)$ and $\mathrm{B}_\mu(u,v)$ are the Heaviside theta function, beta function and incomplete beta function, respectively. Use of these in~(\ref{R1Fapp}) with $F_j(\lambda)$ and $G_k(\lambda)$ given by~(\ref{Fjkev}),~(\ref{Gjkev}) leads to
\begin{eqnarray}
\nonumber
\fl
\mathcal{L}^{-1}\{s^{1-n(n+2\alpha+1)/2}\Phi_{j,k}(s\mu)\}(t)&&\big|_{t=1}= \Gamma(k+\alpha) \mu^{j+\alpha-1} \Big[\frac{2\mu^{k+\alpha}\mathrm{B}(1-k-\alpha,\gamma-j)}{\Gamma(\gamma-j)}\\
\fl
&&- \frac{2\mu^{k+\alpha}\mathrm{B}_{\mu}(1-k-\alpha,\gamma-j)}{\Gamma(\gamma-j)}
 -\frac{(1-\mu)^{\gamma-j}}{\Gamma(\gamma-j+1)}\Big].
\end{eqnarray}
Expressing the (complete) beta function in terms of gamma functions and simplifying gives us~(\ref{R1Fb}). Similarly, when $n$ is odd, using~(\ref{Fjkod}),~(\ref{Gjkod}) we also have
\begin{eqnarray}
\nonumber
\mathcal{L}^{-1}\{s^{1-n(n+2\alpha+1)/2}\Phi_{j,n+1}(s\mu)\}(t)\big|_{t=1}=-\frac{\mu^{j+\alpha-1}(1-\mu)^{\gamma-j}}{\Gamma(\gamma-j+1)},\\
~\mathcal{L}^{-1}\{s^{1-n(n+2\alpha+1)/2}\Phi_{n+1,k}(s\mu)\}(t)\big|_{t=1}=0.
\end{eqnarray}
Therefore, the desired expression~(\ref{R1Fb}) follows.

We should add that in reference~\cite{OSZ2010}, for the square case ($m=n$), the distribution~(\ref{PBH}) has been shown explicitly to correspond to the matrix model~(\ref{rho}) using an approach aided by matrix Dirac delta function~\cite{Zhang2016}. Moreover, in the references~\cite{SZ2004} and~\cite{OSZ2010}, the moments and averages pertaining to the fixed trace ensembles have been deduced by considering the matrix Laplace transforms for the unrestricted trace ensemble and thereby formulating the generating functions. The formalism adopted in these references is equivalent to the current approach of working in the eigenvalue space to derive the correlation function and moments. However, for the rectangular case, the matrix-based approach seems mathematically more involved.

\section{Average HCT entropy in terms of integral involving level density of unrestricted trace ensemble}
\label{AppHCTun}

We begin with the expression~(\ref{avHCT}) for the average entropy to obtain
\begin{eqnarray}
\label{avHCTun}
\nonumber
\langle \mathcal{S}_{\omega}\rangle_\mathrm{BH}=\frac{1}{\omega-1}- \frac{C^{(F)}}{(\omega-1)}&&\int_{0}^{\infty}\cdots\int_{0}^{\infty}\left(\sum_{i=1}^{n}\mu_{i}^{\omega}\right)\delta\left(\sum_{i=1}^{n}\mu_{i}-1\right) \\
&&\times \frac{\Delta^{2}(\{\mu\})}{\Delta_{+}(\{\mu\})}\prod_{j=1}^{n}\mu_{j}^{\alpha}\, d\mu_j
\end{eqnarray}
We now focus on the multidimensional integral part of the above expression and introduce an auxiliary gamma function integral as
\begin{equation}
\fl
\frac{1}{\Gamma(\theta)}\int_0^\infty e^{-r}r^{\theta-1}\int_{0}^{\infty}\cdots\int_{0}^{\infty}\left(\sum_{i=1}^{n}\mu_{i}^{\omega}\right)\delta\left(\sum_{i=1}^{n}\mu_{i}-1\right) 
 \frac{\Delta^{2}(\{\mu\})}{\Delta_{+}(\{\mu\})}\prod_{j=1}^{n}\mu_{j}^{\alpha}\, d\mu_j\cdot dr
\end{equation}
where $\theta$ will be fixed later. Letting $\mu_{j}=\lambda_{j}/r$ and some simplification yields
\begin{equation}
\fl
\frac{1}{\Gamma(\theta)}\int_0^\infty \frac{e^{-r}r^{\theta-1}}{r^{\alpha+\gamma+\omega}}\int_{0}^{\infty}\cdots\int_{0}^{\infty}\left(\sum_{i=1}^{n}\lambda_{i}^{\omega}\right)\delta\left(\sum_{i=1}^{n}\lambda_{i}-r\right) 
 \frac{\Delta^{2}(\{\lambda\})}{\Delta_{+}(\{\lambda\})}\prod_{j=1}^{n}\lambda_{j}^{\alpha}\, d\lambda_j\cdot dr.
\end{equation}
 We now set $\theta=\alpha+\gamma+\omega+1$, and integrate over $r$ first by changing the order of integration. This integration is trivial due to the delta function, and leaves us with
\begin{equation}
\frac{1}{\Gamma(\alpha+\gamma+\omega+1)}\int_{0}^{\infty}\cdots\int_{0}^{\infty}\left(\sum_{i=1}^{n}\lambda_{i}^{\omega}\right)
 \frac{\Delta^{2}(\{\lambda\})}{\Delta_{+}(\{\lambda\})}\prod_{j=1}^{n}\lambda_{j}^{\alpha}e^{-\lambda_j}\, d\lambda_j.
\end{equation}
We substitute this in~(\ref{avHCTun}) and introduce the normalization factor for the unrestricted trace ensemble to obtain the average entropy as an integral involving the unrestricted trace Bures--Hall measure~(\ref{jpd-unres}),
\begin{eqnarray}
\nonumber
\langle \mathcal{S}_{\omega}\rangle_\mathrm{BH}&=&\frac{1}{\omega-1}- \frac{C^{(F)}}{C(\omega-1)\Gamma(\alpha+\gamma+\omega+1)}\\
&&\times\int_{0}^{\infty}\cdots\int_{0}^{\infty}\left(\sum_{i=1}^{n}\lambda_{i}^{\omega}\right) \mathcal{P}(\lambda_1,\dots,\lambda_n)\prod_{j=1}^{n}\, d\lambda_j.
\end{eqnarray}
The ratio $C^{(F)}/C$ of the normalization factors can now be replaced by $\Gamma(\alpha+\gamma+1)$ using~(\ref{normratio}) and the multidimensional integral can be reduced to a single integral involving the level density owing to the symmetry between the eigenvalues and, therefore, gives us the result~(\ref{avHCTb}).

\newpage
\section{Calculation of the average HCT entropy}
\label{AppHCT} 

We first use the level density~(\ref{R1Fb}) for the fixed trace ensemble directly to obtain the average HCT entropy using~(\ref{avHCTa}). This yields
\begin{eqnarray}
\nonumber
\langle \mathcal{S}_{\omega}\rangle_\mathrm{BH}=\frac{1}{\omega-1}-\frac{n!\,C^{(F)}}{\omega-1}\sum_{1\leq j < k \leq N} (-1)^{j+k}\\
\times \left[\int_0^1\mu^\omega\Psi_{j,k}(\mu)d\mu-\int_0^1 \mu^\omega\Psi_{k,j}(\mu)d\mu \right]\Pf[H^{(j,k)}],
\end{eqnarray}
where $\Psi_{j,k}(\mu)$ is given by~(\ref{PsiEv}) and~(\ref{PsiOd}). For~(\ref{PsiEv}), we obtain
\begin{eqnarray}
\nonumber
\fl
\int_0^1\mu^\omega\Psi_{j,k}(\mu)d\mu=\frac{2\Gamma(k+\alpha)\Gamma(1-k-\alpha)}{(j+k+2\alpha+\omega)\Gamma(1-j-k-\alpha+\gamma)}\\
\nonumber
\fl
~~~~~ -\frac{2\Gamma(k+\alpha)}{(j+k+2\alpha+\omega)\Gamma(\gamma-j)}[\mathrm{B}(1-k-\alpha,\gamma-j)-\mathrm{B}(j+\alpha+\omega+1,\gamma-j)]\\
-\frac{\Gamma(k+\alpha)}{\Gamma(\gamma-j+1)}\mathrm{B}(j+\alpha+\omega,\gamma-j+1).
\end{eqnarray}
Here, along with elementary integrals, we used $\int_0^1\mu^a \mathrm{B}_\mu(b,c)\,d\mu={(a+1)^{-1}}[\mathrm{B}(b,c)-\mathrm{B}(a+b+1,c)]$. Now, we write the beta functions in terms of gamma functions and simplify to obtain
\begin{eqnarray}
\label{etaeva}
\nonumber
\fl
\int_0^1\mu^\omega\Psi_{j,k}(\mu)d\mu&=&\frac{2\Gamma(k+\alpha)\Gamma(j+\alpha+\omega+1)}{(j+k+2\alpha+\omega)\Gamma(\alpha+\gamma+\omega+1)}
-\frac{\Gamma(k+\alpha)\Gamma(j+\alpha+\omega)}{\Gamma(\alpha+\gamma+\omega+1)}\\
&=&\left(\frac{j-k+\omega}{j+k+2\alpha+\omega}\right)\frac{\Gamma(j+\alpha+\omega)\Gamma(k+\alpha)}{\Gamma(\alpha+\gamma+\omega+1)}.
\end{eqnarray}
Additionally, when $n$ is odd, using~(\ref{PsiOd}) we get
\begin{eqnarray}
\label{etaoda}
\fl
\int_0^1\mu^\omega\,\Psi_{j,n+1}(\mu)d\mu=-\frac{\Gamma(j+\alpha+\omega)}{\Gamma(\alpha+\gamma+\omega+1)}, ~~~\int_0^1\mu^\omega\,\Psi_{n+1,k}(\mu)d\mu=0. 
\end{eqnarray}
Equation~(\ref{avSomg}) then follows using~(\ref{etaeva}) and (\ref{etaoda}) obtained above and defining $\eta_{j,k}=\Gamma(\alpha+\gamma+\omega+1)\int_0^1\mu^\omega\Psi_{j,k}(\mu)d\mu$.

This result can be also obtained using the equation~(\ref{avHCTb}) which involves integration with the level density of the unrestricted trace ensemble. We have
\begin{eqnarray}
\label{Somgb}
\nonumber
\langle \mathcal{S}_{\omega}\rangle_\mathrm{BH}=\frac{1}{\omega-1}-\frac{n!\,C\, \Gamma(\alpha+\gamma+1)}{(\omega-1)\Gamma(\alpha+\gamma+\omega+1)}\sum_{1\leq j < k \leq N} (-1)^{j+k}\\
\times \left[\int_0^\infty\lambda^\omega\Phi_{j,k}(\lambda)d\lambda-\int_0^\infty \lambda^\omega\Phi_{k,j}(\lambda)d\lambda \right]\Pf[H^{(j,k)}],
\end{eqnarray}
where $\Phi_{j,k}(\lambda)=F_j(\lambda)G_k(\lambda)$. Using~(\ref{Fjkev}) and~(\ref{Gjkev}), we obtain
\begin{eqnarray}
\nonumber
\int_0^\infty \lambda^\omega\Phi_{j,k}(\lambda)d\lambda&=&\Gamma(k+\alpha)\left[\frac{2\Gamma(j+\alpha+\omega+1)}{j+k+2\alpha+\omega}-\Gamma(j+\alpha+\omega)\right]\\
&=&\frac{(j-k+\omega)\,\Gamma(j+\alpha+\omega)\Gamma(k+\alpha)}{j+k+2\alpha+\omega},
\end{eqnarray}
where along with the gamma function integral, we used $\int_0^\infty \lambda^a E_b(\lambda)\,d\lambda=\Gamma(a+1)/(a+b)$.
Also, when $n$ is odd, using~(\ref{Fjkod}) and~(\ref{Gjkod}), we obtain
\begin{eqnarray}
\fl
\int_0^\infty\lambda^\omega\,\Phi_{j,n+1}(\lambda)d\lambda=-\Gamma(j+\alpha+\omega), ~~~\int_0^\infty\lambda^\omega\,\Phi_{n+1,k}(\lambda)d\lambda=0. 
\end{eqnarray}
Substituting these in~(\ref{Somgb}) gives back~(\ref{avSomg}) again. 

\section{Calculation of the average von Neumann entropy}
\label{AppVN} 

We will calculate the average von Neumann entropy by taking the limit $\omega\to 1$ in the average HCT entropy. 
First of all, we notice that, since $\sum_{i=1}^n\mu_i^\omega=1$ for $\omega=1$, the ensemble average over it using the fixed trace jpd also yields $\langle\sum_{i=1}^n\mu_i^\omega\rangle_\mathrm{BH}=\int_{0}^{1}\mu^\omega R_1^{(F)}(\mu)d\mu=1$. Therefore, if we compare the respective factors appearing with $1/(\omega-1)$ in the second term on the right side of~(\ref{avHCTa}) and~(\ref{avSomg}), we obtain 
\begin{equation}
\frac{n!\, C^{(F)}}{\Gamma(\alpha+\gamma+\omega+1)} \sum_{1\leq j<k\leq N} (-1)^{j+k} (\eta_{j,k}-\eta_{k,j} )\Pf \left[H^{(j,k)}\right]\Big|_{\omega=1} =1.
\end{equation}
This, in turn, leads us to the result
\begin{equation}
\label{eq0}
\fl
\sum_{1\leq j<k\leq N}\left[ \frac{2}{N(N-1)}-\frac{n!\, C^{(F)}}{\Gamma(\alpha+\gamma+\omega+1)} (-1)^{j+k} (\eta_{j,k}-\eta_{k,j} )\Pf \left[H^{(j,k)}\right]\right]_{\omega=1} =0.
\end{equation}
Now, we can write down~(\ref{avSomg}) in the following form:
\begin{eqnarray}
\nonumber
\fl \langle \mathcal{S}_{\omega}\rangle_{\mathrm{BH}}= \frac{1}{(\omega-1)} \sum_{1\leq j<k\leq N} \left[\frac{2}{N(N-1)}-\frac{n!\, C^{(F)}(-1)^{j+k} (\eta_{j,k}-\eta_{k,j} ) \Pf \left[H^{(j,k)}\right]}{\Gamma(\alpha+\gamma+\omega+1)}\right]\\
\fl
=\sum_{1\leq j<k\leq N} \left[\frac{2\Gamma(\alpha+\gamma+\omega+1)-n!\, C^{(F)}N(N-1)(-1)^{j+k} (\eta_{j,k}-\eta_{k,j} ) \Pf \left[H^{(j,k)}\right]}{N(N-1)(\omega-1)\Gamma(\alpha+\gamma+\omega+1)}\right].
\end{eqnarray}
In view of the identity~(\ref{eq0}) above, this is clearly of 0/0 form for $\omega=1$. Therefore, we may apply L'Hospital's rule in this expression to evaluate the $\omega\to 1$ limit, and thereby obtain the average von Neumann entropy. We use the following results:
\begin{eqnarray}
\fl
\frac{\partial}{\partial \omega}\Gamma(\alpha+\gamma+\omega+1)\stackrel{\omega\to1}{=}\Gamma(\alpha+\gamma+2)\psi(\alpha+\gamma+2),\\
\fl
\frac{\partial}{\partial \omega}(\omega-1)\Gamma(\alpha+\gamma+\omega+1)\stackrel{\omega\to1}{=}\Gamma(\alpha+\gamma+2),\\
\nonumber
\fl
\label{etasymasym}
\frac{\partial}{\partial \omega}\eta_{j,k}\stackrel{\omega\to1}{=}\frac{2\Gamma(j+\alpha+1)\Gamma(k+\alpha+1)}{(j+k+2\alpha+1)^2}\\
\fl
~~~~~~~ +\frac{(j-k+1)}{(j+k+2\alpha+1)}\Gamma(j+\alpha+1)\Gamma(k+\alpha)\psi(j+\alpha+1);~~~j,k=1,...,n,\\
\fl
\frac{\partial}{\partial \omega}\eta_{j,n+1}\stackrel{\omega\to1}{=}-\Gamma(j+\alpha+1)\psi(j+\alpha+1),~~\frac{\partial}{\partial \omega}\eta_{n+1,k}\stackrel{\omega\to1}{=}0;~~~j,k=1,...,n.
\end{eqnarray}
We note that the first term in~(\ref{etasymasym}) is symmetric in $j$ and $k$, and since we are concerned with the difference $\eta_{j,k}-\eta_{k,j}$, we may drop it and define $\xi_{j,k}$ as in~(\ref{xjkev}) for $j,k=1,...,n$, along with $\xi_{j,n+1}$ and $\xi_{n+1,k}$ given in (\ref{xjnp1}). Therefore, the limit gives us
\begin{eqnarray}
\nonumber
\fl
\sum_{1\leq j<k\leq N} \left[\frac{2\Gamma(\alpha+\gamma+2)\psi(\alpha+\gamma+2)-n!\, C^{(F)}N(N-1)(-1)^{j+k} (\xi_{j,k}-\xi_{k,j} ) \Pf \left[H^{(j,k)}\right]}{N(N-1)\Gamma(\alpha+\gamma+2)}\right],
\end{eqnarray}
which, in turn, leads to~(\ref{avS1}) after separating the two terms in the summation.

Instead of performing the above limit, the average von Neumann entropy calculation can also be performed by directly averaging over $-\sum_i \mu_i\ln\mu_i$ using the level density expression of the fixed trace ensemble or mapping it to an average over the level density of the unrestricted trace level density, similar to what has been done for the average HCT entropy. This again leads to the same expression as~(\ref{avS1}) after some simplification. 



\section*{References}
\begin{thebibliography}{99}

\bibitem{Neumann1927} von Neumann J 1927 Wahrscheinlichkeitstheoretischer Aufbau der Quantenmechanik {\it G\"{o}ttinger Nachrichten} {\bf 1} 245

\bibitem{Haar1961} Ter Haar D 1961 Theory and applications of the density matrix {\it Rep. Prog. Phys.} {\bf 24} 304

\bibitem{BZ2006} Bengtsson I and \.{Z}yczkowski K 2006 {\it Geometry of quantum states: an introduction to quantum entanglement} (Cambridge University Press, Cambridge, England, 2006)

\bibitem{SZ2004} Sommers H-J and \.Zyczkowski K 2004 Statistical properties of random density matrices {\it J. Phys. A: Math. Gen.} {\bf 37} 8457

\bibitem{ZS2001} \.Zyczkowski K and Sommers H J 2001 Induced measures in the space of mixed quantum states {\it J. Phys. A: Math. Gen.} {\bf 34} 7111

\bibitem{ZS2003} \.Zyczkowski K and Sommers H-J 2003 Hilbert--Schmidt volume of the set of mixed quantum states \textit{J. Phys. A: Math. Gen.} {\bf 36 } 10115

\bibitem{Slater1999} Slater P B 1999 Hall normalization constants for the Bures volumes of the n-state quantum systems \textit{J. Phys. A: Math. Gen.} {\bf 32} 8231

\bibitem{BS2001} Byrd M S and Slater P B 2001 Bures measures over the spaces of two- and three-dimensional density matrices {\it Phys. Lett. A} {\bf 283} 152 

\bibitem{SZ2003} Sommers H-J and  \.Zyczkowski K  2003 Bures volume of the set of mixed quantum states \textit{J. Phys. A: Math. Gen.} {\bf 36 } 10083

\bibitem{Brausntein1996} Braunstein S L 1996 Geometry of quantum inference {\it Phys. Lett. A} {\bf 219} 169

\bibitem{Hall1998} Hall M J W 1998 Random quantum correlations and density operator distributions  {\it Phys. Lett. A.} {\bf 242} 123

\bibitem{NC2000} Nielsen M A and Chuang I L 2000 {\it Quantum Computation and Quantum Information} (Cambridge : Cambridge University Press)

\bibitem{ZPNC2011}  \.{Z}yczkowski K, Penson K A, Nechita I and Collins B 2011 Generating random density matrices {\it J. Math. Phys.} {\bf 52} 06220

\bibitem{OSZ2010} Osipov V A, Sommers H J and \.Zyczkowski K  2010 Random Bures mixed states and the distribution of their purity {\it J. Phys. A: Math. Theor.} {\bf43} 055302

\bibitem{Hubner1992} H\"ubner M 1992 Explicit computation of the Bures distance for density matrices \textit{Phys. Lett. A} {\bf163} 239

\bibitem{Uhlmann1976} Uhlmann A 1976 The ``transition probability" in the state space of a*-algebra {\it Rep. Math. Phys.} {\bf 9} 273

\bibitem{Bures1969} Bures D J C 1969  An extension of Kakutani's theorem on infinite product measures to the tensor product of semifinite $ w*$-algebras {\it Trans. Am. Math. Soc.} {\bf 135} 199

\bibitem{Jos1994} Jozsa R 1994 Fidelity for Mixed Quantum States {\it J. Mod. Opt.} {\bf 41} 2315

\bibitem{Fisher1925} Fisher R A 1925 Theory of statistical estimation  \textit{Math. Proc. Camb. Phil. Soc.} {\bf 22} 700 

\bibitem{Lubkin1978} Lubkin E 1978 Entropy of an $n$-system from its correlation with a $k$-reservoir {\it J. Math. Phys.} {\bf 19} 1028

\bibitem{LP1988} Lloyd S and Pagels H 1988 Complexity as thermodynamic depth {\it Ann. Phys., NY} {\bf 188} 186

\bibitem{Page1993} Page D N 1993 Average entropy of a subsystem {\it Phys. Rev. Lett.} {\bf 71} 1291

\bibitem{CSZ2006} Cappellini V, Sommers H J and \.Zyczkowski K 2006 Distribution of $G$ concurrence of random pure states {\it Phys. Rev. A} {\bf 74} 062322

\bibitem{Giraud2007} Giraud O 2007 Purity distribution for bipartite random pure states {\it J. Phys. A: Math.Theor.} {\bf 40} 2793

\bibitem{Sen1996} Sen S 1996 Average entropy of a quantum subsystem {\it Phys. Rev. Lett.} {\bf 77} 1

\bibitem{MBL2008} Majumdar S N, Bohigas O and Lakshminarayan A 2008 Exact minimum eigenvalue distribution of an entangled random pure state {\it J. Stat. Phys.} {\bf 131} 33

\bibitem{FMPPS2008} Facchi P, Marzolino U, Parisi G, Pascazio S and Scardicchio A 2008 Phase transitions of bipartite entanglement {\it Phys. Rev. Lett.} {\bf 101} 050502

\bibitem{KAT2008} Kubotani H, Adachi S and Toda M 2008 Exact formula of the distribution of Schmidt eigenvalues for dynamical formation of entanglement in quantum chaos {\it Phys. Rev. Lett.} {\bf 100} 240501

\bibitem{ATK2009} Adachi S, Toda M and Kubotani H 2009 Random matrix theory of singular values of rectangular complex matrices: I. Exact formula of one-body distribution function in fixed-trace ensemble {\it Ann. Phys.} {\bf 324} 2278

\bibitem{Vivo2010} Vivo P 2010 Entangled random pure states with orthogonal symmetry: exact results {\it J. Phys. A: Math. Theor.} {\bf 43} 405206

\bibitem{PFPPS2010}  De Pasquale A, Facchi P, Parisi G, Pascazio S and Scardicchio A 2010  Phase transitions and metastability in the distribution of the bipartite entanglement of a large quantum system {\it Phys. Rev. A} {\bf81} 052324

\bibitem{CLZ2010} Chen Y, Liu D-Z and Zhou D-S 2010 Smallest eigenvalue distribution of the fixed-trace Laguerre beta-ensemble {\it J. Phys. A: Math. Theor.} {\bf 43} 315303

\bibitem{NMV2010} Nadal C, Majumdar S N and Vergassola M 2010 Phase transitions in the distribution of bipartite entanglement of a random pure state {\it Phys. Rev. Lett.} {\bf 104} 110501

\bibitem{AV2011} Akemann G and Vivo P 2011 Compact smallest eigenvalue expressions in Wishart--Laguerre ensembles with or without a fixed trace {\it J. Stat. Mech.} {\bf 2011} P05020

\bibitem{Majumdar2011} Majumdar S N 2010 Extreme eigenvalues of Wishart matrices: application to entangled bipartite system {\it The Oxford Handbook of Random Matrix Theory} (Oxford: Oxford University Press) (arXiv:1005.4515)

\bibitem{LZ2011} Liu D-Z and Zhou D-S 2011 Local statistical properties of Schmidt eigenvalues of bipartite entanglement for a random pure state  {\it Int. Math. Res. Not.} {\bf 2011} 725

\bibitem{KP2011} Kumar S and Pandey A 2011 Entanglement in random pure states: spectral density and average von Neumann entropy {\it J. Phys. A: Math. Theor.} {\bf 44}  445301

\bibitem{NMV2011} Nadal C, Majumdar S N and Vergassola M 2011 Statistical distribution of quantum entanglement for a random bipartite state {\it J. Stat. Phys.} {\bf 142} 403

\bibitem{BN2012} Borot G and Nadal C 2012 Purity distribution for generalized random Bures mixed states {\it J. Phys A: Math.Theor.} {\bf 45} 075209

\bibitem{BTL2012} Bhosale U T, Tomsovic S and Lakshminarayan A 2012 Entanglement between two subsystems, the Wigner semicircle and extreme-value statistics {\it Phys. Rev. A} {\bf 85} 062331

\bibitem{VPO2016} Vivo P, Pato M P and Oshanin G 2016 Random pure states: Quantifying bipartite entanglement beyond the linear statistics {\it Phys. Rev. E} {\bf 93} 052106

\bibitem{Wei2017} Wei L 2017 Proof of Vivo-Pato-Oshanin's conjecture on the fluctuation of von Neumann entropy {\it Phys. Rev. E} {\bf 96} 022106 

\bibitem{KSA2017} Kumar S, Sambasivam B and Anand S 2017 Smallest eigenvalue density for regular or fixed-trace complex Wishart--Laguerre ensemble and entanglement in coupled kicked tops {\it J. Phys. A: Math. Theor.} {\bf 50} 345201

\bibitem{TLSB2018} Tomsovic S, Lakshminarayan A, Srivastava S C L and B\"{a}cker A 2018 Eigenstate entanglement between quantum chaotic subsystems: Universal transitions and power laws in the entanglement spectrum {\it Phys. Rev. E} {\bf 98} 032209

\bibitem{Wei2018} Wei L 2018 On the exact variance of Tsallis entropy in a random pure state {\it Entropy} {\bf 21} 539

\bibitem{FK2016} Forrester P J and Kieburg M 2016 Relating the Bures measure to the Cauchy two-matrix model {\it Commun. Math. Phys.}  {\bf 342} 151

\bibitem{BGS2009} Bertola M, Gekhtman M and Szmigielski J 2009 The Cauchy two-matrix model {\it Commun. Math. Phys.} {\bf 287} 983

\bibitem{HC1967} Havrda J and Charv\'{a}t F 1967 Quantification method of classification processes. Concept of structural a-entropy {\it Kybernetika} {\bf 3} 30

\bibitem{Tsallis1988} Tsallis C 1988 Possible generalization of Boltzmann-Gibbs statistics {\it J. Stat. Phys.} {\bf 52} 479

\bibitem{Dyson1972} Dyson F J 1972 A Class of Matrix Ensembles {\it J. Math. Phys.} {\bf 13} 90

\bibitem{Mehta2004} Mehta M L 2004 {\it Random Matrices} (New York: Academic)

\bibitem{Forrester2010} Forrester P J 2010 \textit{Log-Gases and Random Matrices} (Princeton, NJ: Princeton University Press)

\bibitem{Kostov1989} Kostov I K 1989 $O(n)$ vector model on a planar random lattice: spectrum of anomalous dimensions {\it Mod. Phys. Lett. A} {\bf 4} 217

\bibitem{EK1995} Eynard B and Kristjansen C 1995 Exact solution of the $O(n)$ model on a random lattice { \it Nucl. Phys. B} {\bf 455} 577

\bibitem{FL2018} Forrester P J and Li S-H 2018 Fox H-kernel and $\theta$-deformation of the Cauchy two-matrix model and Bures ensemble {\it arXiv:1811.03183}

\bibitem{HL2017} Hu X-B and Li S-H 2017 The partition function of the Bures ensemble as the $\tau$-function of BKP and DKP hierarchies: continuous and discrete {\it J. Phys. A: Math. Theor.} {\bf 50} 285201

\bibitem{Dyson1970} Dyson F J 1970 Correlations between eigenvalues of a random matrix {\it J. Math. Phys.} {\bf 19} 235

\bibitem{Mehta1971} Mehta M L 1971 A note on correlations between eigenvalues of a random matrix {\it Commun. Math. Phys.} {\bf 20} 245

\bibitem{Mehta1976} Mehta M L 1976 A note on certain multiple integrals {\it J. Math. Phys.} {\bf 17} 2198

\bibitem{BN1991} Br\'{e}zin E and Neuberger H 1991 Multicritical points of unoriented random surfaces {\it Nucl. Phys. B.} {\bf 350} 513.

\bibitem{Nagao2007} Nagao T 2007 Pfaffian expressions for random matrix correlation functions {\it J. Stat. Phys.} {\bf 129} 1137.

\bibitem{AK2007} Akemann G and Kanzieper E 2007 Integrable structure of Ginibre's ensemble of real random matrices and a Pfaffian integration theorem {\it J. Stat. Phys.} {\bf 129} 1159

\bibitem{KG2010} Kieburg M and Guhr G 2010 A new approach to derive Pfaffian structures for random matrix ensembles {\it J. Phys. A: Math. Theor.} {\bf 43} 135204

\bibitem{PM1983} Pandey A and Mehta M L 1983 Gaussian ensembles of random hermitian matrices intermediate between orthogonal and unitary ones {\it Commun. Math. Phys.} {\bf 87} 449

\bibitem{MP1983} Mehta M L and Pandey A 1983 On some Gaussian ensembles of Hermitian matrices {\it J. Phys. A: Math. Gen.} {\bf 16} 2655

\bibitem{PS1991} Pandey A and Shukla P 1991 Eigenvalue correlations in the circular ensembles {\it J. Phys. A: Math. Gen.} {\bf 24} 3907

\bibitem{FNH1999} Forrester P J, Nagao T and Honner G 1999 Correlations for the orthogonal-unitary and symplectic-unitary transitions at the hard and soft edges {\it Nucl. Phys. B} {\bf 553} 601

\bibitem{NF2003} Nagao T and Forrester P J 2003 Dynamical correlations for circular ensembles of random matrices {\it Nucl. Phys. B} {\bf 660} 557

\bibitem{KP2009} Kumar S and Pandey A 2009 Universal spectral correlations in orthogonal-unitary and symplectic-unitary crossover ensembles of random matrices {\it Phys. Rev. E} {\bf 79} 026211

\bibitem{KP2011b} Kumar S and Pandey A 2011 Crossover ensembles of random matrices and skew-orthogonal polynomials {\it Ann. Phys.} {\bf 326} 1877

\bibitem{Schur1911} Schur I 1911 \"{U}ber die Darstellung der symmetrischen und der alternirenden Gruppe durch gebrochene lineare Substitutuionen, {\it J. Reine Angew. Math.} {\bf 139} 155

\bibitem{IOTZ1995} Ishikawa M, Okanda S, Tagawa H and Zeng J 1995 Generalizations of Cauchy's determinant and Schur's Pfaffians, {\it Linear and Multilinear Algebra} {\bf 39} 251

\bibitem{Bruijn1955} de Bruijn N G 1955 On some multiple integrals involving determinants {\it J.Indian Math.Soc.} {\bf 19} 133

\bibitem{Kieburg2012} Kieburg M 2012 Mixing of orthogonal and skew-orthogonal polynomials and its relation to Wilson RMT \textit{J. Phys. A: Math. Theor.} {\bf 45} 205203

\bibitem{Okada2017} Okada S 2017 Pfaffian Formulas and Schur Q-Function Identities {\it arXiv:1706.01029}

\bibitem{Hoskins2009} Hoskins R F 2009 {\it Delta functions: Introduction to generalised functions} (Cambridge, UK: Woodhead Publishing Limited)

\bibitem{Renyi1961} R\'enyi A 1961 On measures of entropy and information {\it Proc. of the Fourth Berkeley Symp. on Mathematical Statistics and Probability} vol 1
pp 547--561

\bibitem{Zhang2016} Zhang L 2016 Dirac delta function of matrix argument {\it arXiv:1607.02871}

\end{thebibliography}
\end{document}